**Title**

# Manifold-based transformation of probability distributions: application to the inverse problem of reconstructing distributions from experimental data

Tomotaka Oroguchi[1,2*], Rintaro Inoue[3], and Masaaki Sugiyama[3]

[1]Department of Physics, Faculty of Science and Technology, Keio University, 3-14-1 Hiyoshi, Kohoku-ku, Yokohama 223-8522, Japan.

[2]RIKEN SPring-8 Center, 1-1-1 Kouto, Sayo-cho, Sayo-gun, Hyogo 679-5148, Japan

[3]Institute for Integrated Radiation and Nuclear Science, Kyoto University, Kumatori, Sennan-gun, Osaka 590-0494, Japan.

*Corresponding author

oroguchi@phys.keio.ac.jp

**Abstract**

Information geometry is a mathematical framework that elucidates the manifold structure of the probability distribution space (**p**-space), providing a systematic approach to transforming probability distributions (PDs). In this study, we utilized information geometry to address the inverse problems associated with reconstructing PDs from experimental data. Our initial finding is that the Kullback–Leibler divergence, often considered non-metric owing to its asymmetry, can serve as a valid metric under specific geometric conditions on the manifold. Based on this finding, we formulated the manifold-based gradient descent (MBGD) method, which was employed to visualize the internal structures—represented as PDs—of two types of systems: those with static constituent elements and those with dynamic state transitions. Through the application of MBGD, we successfully reconstructed the underlying PDs for both types of systems, outperforming the standard gradient descent methods that neglect the manifold structure of **p**-space. Therefore, the present results demonstrate the essentiality of accounting for the manifold structure of **p**-space in the inverse problems of reconstructing PDs. The ability of MBGD to accurately reconstruct PDs for systems with dynamic state transitions underscores its potential to provide valuable physical insights by visualizing internal structures.

# I. INTRODUCTION

Information geometry [1] is a framework that reveals the manifold structure inherent in the probability distribution space (**p**-space). According to Chentsov's invariance theorem [2], the Fisher information metric is the unique Riemannian metric on **p**-space [3,4]. Nagaoka and Amari further demonstrated that this metric endows **p**-space with a dually affine manifold structure [5,6]. Naturally, this manifold structure results in the Kullback-Leibler (KL) divergence [7], serving as a canonical (albeit asymmetric) divergence between probability distributions (PDs) [8,9]. Since these fundamental mathematical principles define how PDs evolve or should be transformed, i.e., how they move on **p**-space, information geometry is now recognized as a fundamental tool not only in information science [10,11] but also in various branches of physics [12], including phase transitions [13-16], complexity [17-19], non-equilibrium statistical mechanics [20-22], and quantum mechanics [23-27].

Despite the invaluable utility of information geometry, its application to inverse problems involving the reconstruction of PDs from experimental data remains largely unexplored. In such problems, the PDs representing the state distributions of system components are not directly observable and must be inferred from indirect measurements. These inverse problems hold significant implications across various physical domains [28-42]. Systems in these applications can be broadly categorized into two types: (i) those whose constituent elements remain static and (ii) those whose elements undergo state transitions. Systems falling into the former category are of particular interest in engineering applications, such as shape characterization of synthesized nanoparticles [28,29]. From a physics perspective, type (ii) systems pose a greater challenge and hold more significance. In these systems, reconstructing PDs requires an understanding of the physical properties of the system elements, whereas the resulting outcomes contribute to refining this understanding. Such applications are in high demand in soft matter physics [30-35].

The inverse problem of reconstructing a PD is formulated as follows: Consider a discretized system with $n$ states, where a model PD is represented as $\mathbf{p} = (p_1, \ldots, p_n)$, and the observables for the system are denoted as $\mathbf{Y} = (y(1), \ldots, y(m))$ with $m$ data points. Let $\mathbf{Y}(\mathbf{p})$ be the calculated observables based on $\mathbf{p}$, and let $\mathbf{Y}^{\mathrm{EXP}}$ represent the experimental data. We define an objective functional as $\Phi(\mathbf{Y}(\mathbf{p})\|\mathbf{Y}^{\mathrm{EXP}})$, which evaluates the discrepancy between the model and the experimental data (hereafter, designated as $\Phi(\mathbf{p})$). The goal of the inverse problem is to find the optimal $\mathbf{p}$ by minimizing $\Phi(\mathbf{p})$ (Fig. 1(a)). The key task in this inverse problem lies in the transformation of $\mathbf{p}$ during minimization, which must adhere to two constraints: the normalization condition of PD that is, $\sum_{i=1}^{n} p_i = 1$, and the non-negativity of the probability values. These constraints render the minimization process unstable, making the direct reconstruction of PDs from experimental data using the objective functional $\Phi(\mathbf{p})$ a difficult task. Therefore, various domain-specific algorithms have been developed for different applications. However, most approaches require parameter tuning by domain specialists. To achieve wide applicability, a universal approach based on a mathematical understanding of $\mathbf{p}$-space is essential.

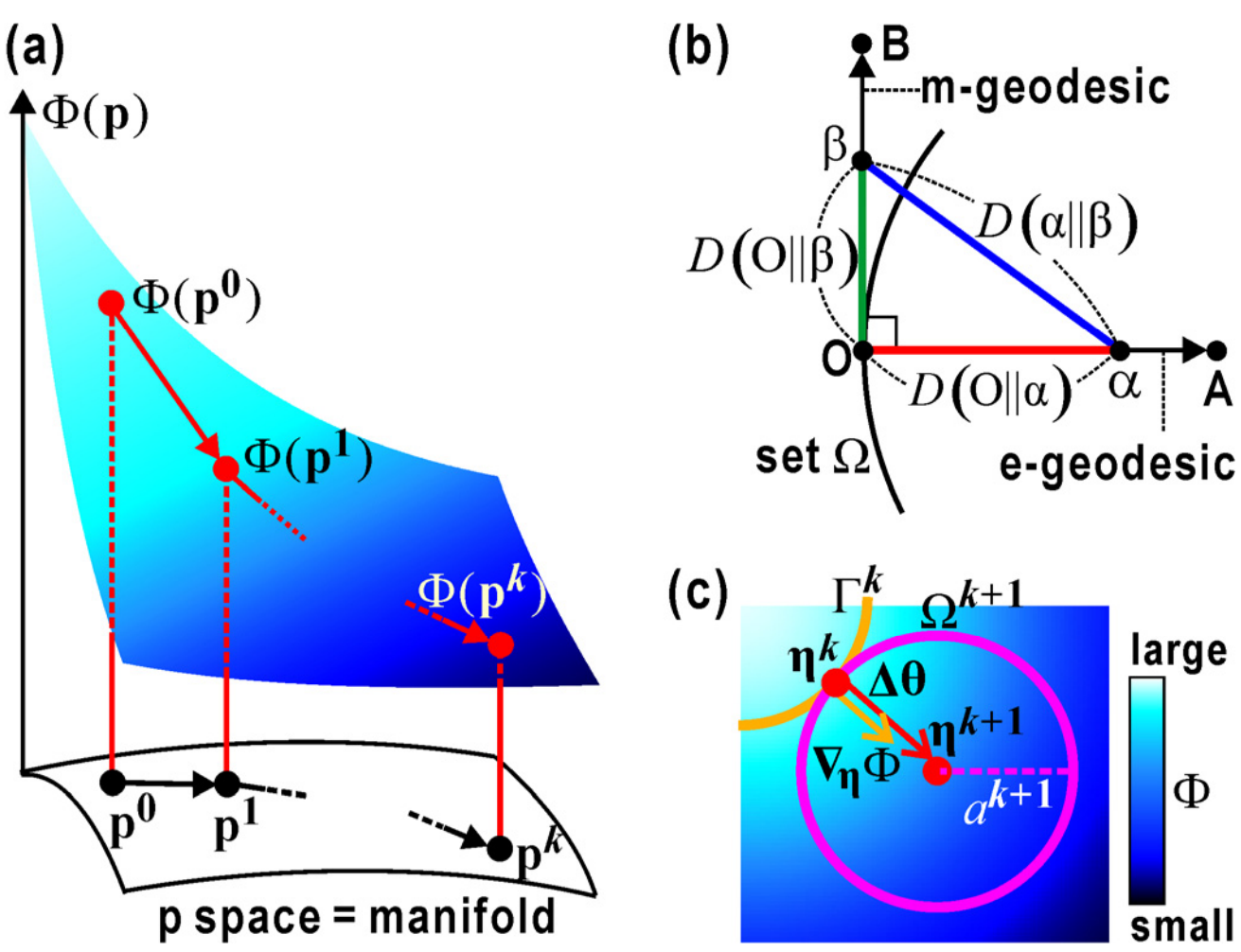

**FIG. 1.** Schematic of the inverse problem reconstructing of the probability distribution $\mathbf{p}$ from experimental data. (a) Iterative transformation of $\mathbf{p}$ from $\mathbf{p}^0$ under an objective functional $\Phi(\mathbf{p})$. (b) Geometry of the transformation path from point O to A in the $\mathbf{p}$-space. (c) Geometry of the gradient descent in the $\mathbf{p}$ space.]

Based on insights from information geometry, the constraints for PD, which mathematically correspond to Chentsov's invariance, transform **p**-space into a manifold. Therefore, if transformations of PDs during the minimization process can be conducted to follow this manifold (Fig. 1(a)), the constraints will be automatically satisfied. Such manifold-based transformations would stabilize the minimization process and thereby enable the reconstruction of true PDs. However, the exact form of such transformations remains unclear, as defining a physical quantity to act as a distance on **p**-space poses a challenge. In this study, we found that the KL divergence, typically considered non-metric owing to its asymmetry, can satisfy the properties of a distance under specific geometric conditions. Such condition corresponds to an infinitesimal displacement along the exponential geodesic (e-geodesic) connecting two PDs (Fig. 1(b)). This finding suggests that in the vicinity of a point on **p**-space, the manifold can be locally approximated as a Euclidean space. This approximate Euclidean space allows for the formulation of the gradient descent on **p**-space (Fig. 1(c)) as follows:

$$\Delta \log \mathbf{p} = -\tau \nabla_{\mathbf{p}} \Phi\left(\mathbf{p}\right) + C\,, \tag{1}$$

where log **p** = (log $p_1$, …, log $p_n$), $C = -\log\left(\sum_{i=1}^{n} p_i \exp\left(-\tau\, \partial\Phi / \partial p_i\right)\right)$ and $\tau$ denotes the step size. Unlike standard gradient descent methods, the left-hand side of Eq. (1) does not correspond to the Euclidean displacement $\Delta\mathbf{p}$, but rather to the displacement along e-geodesic, $\Delta\log\mathbf{p}$. This signifies how PDs should be transformed on the manifold. Hereafter, we refer to Eq. (1) as a manifold-based gradient descent (MBGD) method.

To validate our findings on the KL divergence and the MBGD method built upon it, we conducted reconstruction simulations using model systems. When developing algorithms for inverse problems, it is necessary to evaluate not only their effectiveness in minimizing the objective functional but also their accuracy in the reconstruction of the underlying distribution that generates the experimental data [29,35,43,44]. However, in practice, the true distribution is not directly observable in general, making it challenging to assess the reconstruction accuracy using real data. Consequently, validation studies for algorithms often employ simulations using pseudo-experimental data generated from a synthesized pseudo-true model system. In this study, we used a particle size distribution (PSD) in solution as the model system, which corresponds to a type (i) system. We performed simulations in which the PSDs were reconstructed from small-angle X-ray scattering (SAXS) data (Fig. 3(a)). This model system allowed for a comparison between the standard gradient descent and MBGD in terms of convergence behavior and accuracy in reconstructing the true distribution. The outcomes of our study unequivocally support our theoretical findings on the KL divergence, underscoring the significance of incorporating the manifold structure of **p**-space in solving inverse problems. Furthermore, we delved into a method for assessing the inherent ill-posedness in inverse problems.

Based on the results obtained from the PSD model system, we applied the MBGD method to a type (ii) system in which the constituent elements undergo state transitions. Protein conformational ensembles were selected as an example of such soft matter systems. Most proteins adopt multiple conformational states, and the equilibrium between these states, that is, the conformational ensemble, is crucial for their biological functions [45-51]. While molecular dynamics (MD) simulations are commonly utilized to investigate these ensembles [52], the accuracy of MD is limited by the approximate nature of the molecular force fields [53]. Consequently, inverse problem approaches that refine MD-derived ensembles using experimental data are gaining importance [32,33,54]. The results presented in this study demonstrate that applying MBGD to such inverse problems provides a powerful framework for visualizing the components of soft matter systems.

## II. THEORY

### A. Proof of the geometric condition under which KL-divergence serves as a distance

Here, we define two nearby points O and A in the **p**-space (or on the manifold), and consider the transformation from $\mathbf{p}^{\mathrm{O}}$ to $\mathbf{p}^{\mathrm{A}}$ (Fig. 1b). The superscripts represent the points on the manifold hereinafter. Our primary objective here is to identify a local coordinate system and transformation path that satisfy the geometric conditions of the Euclidean space.

(i) An affine coordinate system can be established.

(ii) For any point α on the path, the distance from point O, $d\left(\mathrm{O}\,\|\,\alpha\right)$, can be defined, satisfying the

following conditions:

(ii-a) Positivity and symmetry: $d\left(\mathrm{O}\|\,\alpha\right)=d\left(\alpha\|\,\mathrm{O}\right)\geq 0$.

(ii-b) Pythagorean theorem: $d\left(\mathrm{O}\|\,\alpha\right)+d\left(\mathrm{O}\|\,\beta\right)=d\left(\alpha\|\,\beta\right)$.

(ii-c) Triangle inequality: $\sqrt{d\left(\mathrm{O}\|\,\alpha\right)}+\sqrt{d\left(\mathrm{O}\|\,\beta\right)}\geq\sqrt{d\left(\alpha\|\,\beta\right)}$.

The information geometry reveals that the manifold structure of **p**-space is a dually flat space [5,6], characterized by the Fisher information metric [3,4]:

$$g_n\left(\frac{\partial}{\partial\mu_j},\frac{\partial}{\partial\mu_k}\right)=\sum_{i=1}^{n}p_i\left(\frac{\partial}{\partial\mu_j}\log p_i\right)\left(\frac{\partial}{\partial\mu_k}\log p_i\right). \tag{2}$$

where $\partial/\partial\mu_j$ denotes the tangent vector regarding the local coordinate system $\boldsymbol{\mu}$. In the dually flat manifold, mutually dual affine coordinate systems exist, $\boldsymbol{\theta}=(\theta_1,\ \ldots,\ \theta_{n-1})$ and $\boldsymbol{\eta}=(\eta_1,\ \ldots,\ \eta_{n-1})$, expressed as follows:

$$\theta_i=\log\left(p_i/p_n\right)\quad\text{and}\quad\eta_i=p_i. \tag{3}$$

$\boldsymbol{\theta}$ and $\boldsymbol{\eta}$ are called exponential-connection (e-connection) and mixture-connection (m-connection) coordinate systems, respectively. In these coordinate systems, the Fisher information metric becomes

$$g_n\left(\frac{\partial}{\partial\theta_i},\frac{\partial}{\partial\eta_j}\right)=\delta_{ij}. \tag{4}$$

Since only $\boldsymbol{\theta}$ and $\boldsymbol{\eta}$ are dual affine coordinate systems on the manifold of **p**-space, the use of either of them is necessary for the Euclidean space condition (i). Here, $\boldsymbol{\eta}$ is selected as a coordinate system, resulting in the introduction of the KL divergence [7,55,56], which is defined as follows:

$$D\left(\mathrm{O}\|\,\alpha\right)=\sum_{i=1}^{n}p_i^{\mathrm{O}}\,\log\frac{p_i^{\mathrm{O}}}{p_i^{\alpha}}, \tag{5}$$

for the two points O and α. $D$ is non-negative and quantifies the difference between the two distributions $\mathbf{p}^{\mathrm{O}}$ and $\mathbf{p}^{\mathrm{A}}$.

Here, we assume that point α lies on the e-geodesic [57] connecting points O and A (Fig. 1(b)).

Since e-geodesic is an affine line with respect to $\boldsymbol{\theta}$, the point $\alpha$ and tangent vector $\boldsymbol{\upsilon}$ along this line can be expressed as

$$\theta_i^{\alpha} = \theta_i^{\mathrm{O}} + s\left(\theta_i^{\mathrm{A}} - \theta_i^{\mathrm{O}}\right) \quad \left(0 \le s \le 1\right) \qquad \text{and} \quad \boldsymbol{\upsilon} = \sum_{i=1}^{n-1}\left(\theta_i^{\mathrm{A}} - \theta_i^{\mathrm{O}}\right)\frac{\partial}{\partial \theta_i}, \tag{6}$$

respectively. For the Euclidean space condition (i), the transformation along an e-geodesic is essential. Furthermore, we consider the displacement from O to A, $\Delta \mathbf{p}^{\mathrm{A}} = \mathbf{p}^{\mathrm{A}} - \mathbf{p}^{\mathrm{O}}$, to be infinitesimal, enabling the disregard of third- or higher-order terms. Consequently, the KL divergence $D\left(\mathrm{O} \| \alpha\right)$ approximately functions as a distance metric for any point $\alpha$ on the e-geodesic between O and A. This point is demonstrated below by proving distance condition (ii).

Since $\sum_{i=1}^{n} \Delta p_i^{\alpha} = 0$ owing to the normalization condition, the second-order approximation of the Taylor expansion of the KL divergence (refer to Appendix A) becomes

$$D\left(\mathrm{O} \| \alpha\right) \simeq D\left(\alpha \| \mathrm{O}\right) \simeq \sum_{i=1}^{n} \frac{\left(\Delta p_i^{\alpha}\right)^2}{2 p_i^{\mathrm{O}}} > 0, \tag{7}$$

thereby satisfying condition (ii-a) [58]. Furthermore, a point $\beta$ is introduced, located on the geodesic mixture [9] (m-geodesic) connecting points O and B (Fig. 1(b)). The m-geodesic is an affine line with respect to $\boldsymbol{\eta}$. This indicates that point $\beta$ and tangent vector $\boldsymbol{\omega}$ along this line can be expressed as

$$\eta_i^{\beta} = \eta_i^{\mathrm{O}} + t\left(\eta_i^{\mathrm{B}} - \eta_i^{\mathrm{O}}\right) \quad \left(0 \le t \le 1\right) \quad \text{and} \quad \boldsymbol{\omega} = \sum_{i=1}^{n-1}\left(\eta_j^{\mathrm{B}} - \eta_j^{\mathrm{O}}\right)\partial/\partial \eta_j, \tag{8}$$

respectively. When the e- and m-geodesics are orthogonal at point O, the inner product of the tangent vectors $\boldsymbol{\upsilon}$ and $\boldsymbol{\omega}$, $g_n\left(\boldsymbol{\upsilon}, \boldsymbol{\omega}\right)$, becomes 0. Therefore, by substituting Eqs. (2–8) into the aforementioned orthogonal relation, the Pythagorean theorem (ii-b) can be obtained [57]:

$$D\left(\alpha \| \beta\right) = D\left(\mathrm{O} \| \alpha\right) + D\left(\mathrm{O} \| \beta\right) \tag{9}$$

(Appendix B). When the e- and m-geodesics are not orthogonal, the sum of the square roots of the KL divergence becomes

$$\sqrt{D(\mathrm{O}\|\alpha)}+\sqrt{D(\mathrm{O}\|\beta)} \simeq \sqrt{\sum_{i=1}^{n}\frac{\left(\Delta p_i^{\alpha}\right)^2}{2p_i^{\mathrm{O}}}}+\sqrt{\sum_{i=1}^{n}\frac{\left(\Delta p_i^{\beta}\right)^2}{2p_i^{\mathrm{O}}}}$$
$$\geq \sqrt{\sum_{i=1}^{n}\frac{\left(\Delta p_i^{\alpha}-\Delta p_i^{\beta}\right)^2}{2p_i^{\mathrm{O}}}} \simeq \sqrt{D(\alpha\|\beta)}, \tag{10}$$

satisfying triangle inequality (ii-c). Based on Eqs. (7), (9), and (10), any infinitesimal change along the e-geodesic connecting two points satisfies Euclidean distance conditions (ii).

Next, the reverse relationship is observed. When the Euclidean space conditions are satisfied on the line connecting points O and A, the line becomes e-geodesic. Consider a set of points, denoted as set Ω, whose KL divergences from point A have the same value with $D(\mathrm{O}\|\mathrm{A})$. Since the Euclidean nature is maintained between points O and A, with point O being a member of set Ω, both the set and point A hold Euclidean properties. In Euclidean space, a set of points equidistant from the center forms a hypersphere, with the normal vector on the surface pointing toward the center. Analogously, set Ω also forms a hypersphere, with any normal vector on the surface pointing toward the center, point A. Therefore, the $i^{\text{th}}$ component of the line connecting points O and A becomes

$$\left\{-\nabla_{\boldsymbol{\eta}} D(\mathrm{O}\|\mathrm{A})\right\}_i=-\frac{\partial D(\mathrm{O}\|\mathrm{A})}{\partial \eta_i}=\theta_i^{\mathrm{A}}-\theta_i^{\mathrm{O}}. \tag{11}$$

A comparison between Eqs. (11) and (6) reveals that the line is e-geodesic.

### B. Formulation of MBGD

We formulated the gradient descent method in **p**-space (MBGD) based on the geometric condition under which the KL divergence serves as a metric. In this formulation, we assume that the geometry of the gradient descent path in a general Euclidean space (Appendix C) holds true between two points on the approximate Euclidean path. We utilize the $\boldsymbol{\eta}$ coordinate system to describe the **p**-space, with $\boldsymbol{\eta}^k$ and $\boldsymbol{\eta}^{k+1}$ representing the model PDs after the $k^{\text{th}}$ and $k+1^{\text{th}}$ transformations through the gradient descent, respectively (Fig. 1(c)). Furthermore, let $\Gamma^k$ be the contour of $\Phi(\boldsymbol{\eta})$ passing through $\boldsymbol{\eta}^k$ and let $\Omega^{k+1}$ be the hypersphere centered at $\boldsymbol{\eta}^{k+1}$ with radius $a^{k+1}$, a parameter that needs to be determined. As demonstrated in the gradient descent path within the general Euclidean space (refer to Appendix C), the relationship between two points on the path can be described by the geometric connection between contour $\Gamma^k$ and hypersphere $\Omega^{k+1}$, which are tangents at point $\boldsymbol{\eta}^k$. In other words, $\boldsymbol{\eta}^k$ represents the extremum at $\Omega^{k+1}$ (Fig. 1(c)). Analogously, the gradient descent for **p**-space can be determined by identifying the extremum of the Lagrangian expressed as follows:

$$L\left(\boldsymbol{\eta}^k\right)=\Phi\left(\boldsymbol{\eta}^k\right)+\left(a^{k+1}-D\left(k+1\| k\right)\right)/\tau . \tag{12}$$

The extremum can be determined by solving $\nabla_{\boldsymbol{\eta}^k} L\left(\boldsymbol{\eta}^k\right)=0$ using Eq. (12), resulting in

$$\Delta\boldsymbol{\theta}=-\tau\nabla_{\boldsymbol{\eta}}\Phi . \tag{13}$$

One advantage of Eq. (13) is its simplicity in determining the normalization constant. This can be shown by rewriting Eq. (13) in the form of Eq. (1). From Eq. (3), each component of Eq. (13) becomes

$$\Delta\log p_i+\tau\frac{\partial\Phi}{\partial p_i}=\Delta\log p_n+\tau\frac{\partial\Phi}{\partial p_n} . \tag{14}$$

Eq. (14) is valid for all $1\le i\le n-1$. Therefore, the right-hand side can be regarded as the normalization constant *C*. Consequently, the normalization condition following the transformation establishes the formulation for *C*, which simplifies Eq. (14) to Eq. (1).

### C. Ill-posedness of the objective functional

In addition to the difficulties associated with the treatment of PDs as reconstruction targets, a typical challenge in inverse problems is their ill-posed nature. This issue arises when the uniqueness of the minimizer of the objective functional is compromised. This uniqueness can be assessed by examining the shape of the objective functional [59-61], whether nonconvex or convex, as shown in Fig. 2. In the case of a non-convex objective functional, multiple minima may exist, making it challenging to determine the minimum corresponding to the true PD (Fig. 2(a)). This situation results in an ill-posed inverse problem, where the solution to which minimization converges depends on the initialization point in **p**-space, that is, the initial PD model.

In contrast, a convex functional, characterized by a positive definite or semi-definite Hessian, has a unique minimum where the true PD may be located (Fig. 2(b)). In this case, regardless of the initial model utilized, the minimization process is likely to converge to this unique minimum. However, the accuracy of the resulting optimal PD may still be influenced by the shape of the minimum, whether flat or sharp. In the case of a flat minimum (upper panel in Fig. 2(b)), the true PD would be surrounded by other solutions with the same objective functional value, making it difficult to determine which one corresponds to the true PD. By contrast, in the case of a strictly sharp minimum (lower panel in Fig. 2(b)), where the Hessian is positive definite, the minimization process converges to a unique PD. Therefore, the sharpness of the minimum in a convex functional can serve as an indicator of the degree of ill-posedness of the inverse problem. In practice, this sharpness is typically characterized by the number of nonzero eigenvalues of the Hessian [61].

Here, we show that the objective functional $\Phi(\mathbf{p})$ is convex when the experimental data are given as expectation values of observables with respect to the PD. In this case, the observables derived from the PD model are expressed as:

$$\mathbf{Y}(\mathbf{p}) = \sum_{i=1}^{n} p_i \mathbf{Y}_i \tag{15}$$

where $\mathbf{Y}_i$ represents the observables for the $i^{\text{th}}$ state. Since experimental errors are typically Gaussian, the negative log-likelihood of a model $\mathbf{p}$ given experimental data $\mathbf{Y}^{\text{EXP}}$ reduces to $\chi^2$ functions:

$$\chi^2 = \frac{1}{M} \sum_{k=1}^{M} \frac{\left( \sum_{i=1}^{N} p_i y_i(k) - y^{\text{EXP}}(k) \right)^2}{\left( \sigma^{\text{EXP}}(k) \right)^2}, \tag{16}$$

where $\sigma^{\text{EXP}}(j)$ denotes the error of the $j^{\text{th}}$ observable. This function is widely utilized to evaluate the disparity between models and experimental data [62,63]. Utilizing $\chi^2$ as the objective functional $\Phi(\mathbf{p})$ results in a Hessian expressed as follows:

$$(\mathbf{H})_{ij} = \frac{\partial^2 \Phi(\mathbf{p})}{\partial p_i \partial p_j} = \frac{1}{2M} \sum_{k=1}^{M} \frac{y_i(k) y_j(k)}{\left\{ \sigma^{\text{EXP}}(k) \right\}^2}. \tag{17}$$

Eq. (17) shows that the Hessian takes the form of a Gram matrix, which is positive definite, or at least positive semi-definite, at any point in **p**-space. Therefore, the convexity of the objective functional is ensured in the present case. Then, the remaining issue concerns the sharpness of the minimum of the objective functional. To address this point, in subsequent reconstruction simulations using the model systems, we investigated the change in sharpness with decreasing reconstruction accuracy. The sharpness of the objective functional was assessed through the number of nonzero eigenvalues, $\Lambda_i$ ($i = 1, \ldots, n$), of its Hessian.

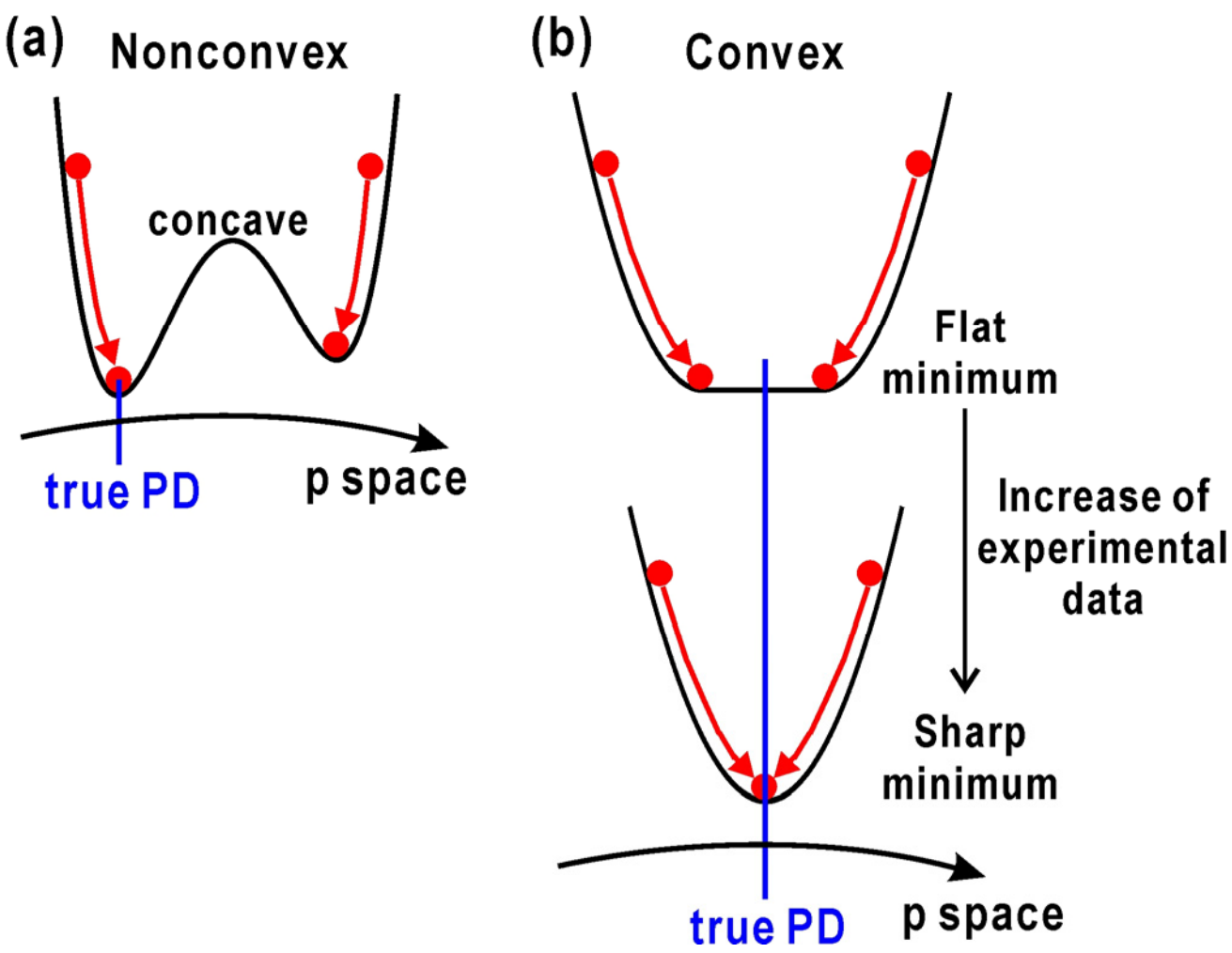

**FIG. 2.** Schematic of the relation between the uniqueness of the minimum and shape of the objective functional. (a) Nonconvex functional. (b) Convex functional: flat (upper) and sharp (lower) minima.

## III. RESULTS

When developing algorithms for inverse problems, it is essential to evaluate their accuracy in reconstructing the underlying distribution that generates the experimental data [29,35,43,44]. To validate our findings on the KL divergence and MBGD method based on it, we conducted simulations using pseudo-experimental data generated from a synthesized pseudo-true system. As such a pseudo-true system, we employed a PSD in solution (Fig. 3(a)), which corresponds to a type (i) system (Subsection A). Reconstructing PSDs is a well-known inverse problem [28,29], and SAXS data has been proven to be effective for such problems through studies utilizing Monte Carlo simulation approaches [29]. Furthermore, we investigated the relationship between the ill-posedness of the inverse problem and sharpness of the objective functional using the PSD system (Subsection B).

Then, as an application example of MBGD to a type (ii) system – where constituent elements undergo state transitions – we addressed the inverse problem of reconstructing protein conformational ensembles from SAXS data. Building on the simulation results from the PSD system, we initially assessed the feasibility of the approach using a pseudo-true ensemble generated by coarse-grained MD (CGMD) simulations [64] (Subsection C). Finally, we applied the MBGD method to the experimental SAXS data for *Streptacidiphilus jiangxiensis* glucosamine kinase (SjGlcNK) [65], which undergoes significant conformational changes upon substrate binding, utilizing both all-atom MD (AAMD) and CGMD simulations (Subsection D).

## A. Reconstruction simulation of particle size distribution from SAXS data

First, we outline the setup for the PSD reconstruction simulations. In these simulations, we assumed a dilute solution in which inter-particle interference could be neglected. For a pseudo-true PSD, $p^{\text{true}}(R)$, which represents a PD over the particle radius $R$, we utilized a model comprising a mixture of two Gaussian functions (Fig. 3(b)). While this Gaussian mixture model may not be realistic as a PSD [29], it allows for a more straightforward interpretation of the validation simulation results (Figs. 4(f) and 5(c)). The $p^{\text{true}}(R)$ was discretized by grouping particles within each 0.2 Å bin along $R$ into a single-size state ($n$ = 1,100 states). The SAXS data $I_i(Q)$ for the $i^{\text{th}}$ particle size $R_i$, were calculated using the scattering function of a sphere, $I_i(Q)=\left[\Delta\rho j_1(QR_i)/R_i^3\right]^2$ ($Q = 4\pi\sin\theta/\lambda$, where $\lambda$ and $2\theta$ represent the wavelength of the incident X-ray beam and scattering angle, respectively) (Fig. 3(c)). $\Delta\rho$ represents the electron density contrast between the particle and solvent, and was set as 1 for simplicity. Subsequently, the pseudo-experimental SAXS data, $I^{\text{EXP}}(Q)$, were determined as expectation values with respect to $p^{\text{true}}(R)$, and thus expressed as $I^{\text{EXP}}(Q)=\sum_{i=1}^{n}p^{\text{true}}(R_i)I_i(Q)$. Furthermore, the SAXS data calculated from $p(R)$ during the minimization process are expressed as $I^{\text{CALC}}(Q)=\sum_{i=1}^{n}p(R_i)I_i(Q)$. For theoretical simplicity, noise was not introduced into $I^{\mathrm{EXP}}(Q)$, resulting in a complete minimization yielding an $\chi^2$ value of 0 (Eq. (16)). In the subsequent data plots, dependencies on both $Q$ and $QR_g/4\pi$ are represented, where $R_g$ represents the radius of gyration of the particle estimated from $I^{\mathrm{EXP}}(Q)$ using Guinier approximation [66], allowing for discussions independent of the particle size. For the simulations, we utilized SAXS data up to $Q$ = 0.5 Å$^{-1}$ ($QR_g/4\pi$ = 2.69), which aligns with the typical upper limit of $Q$ accessible in standard SAXS experiments. We adopted a uniform distribution as the initial PSD $p^0(R)$. The accuracy of $p(R)$ was assessed using the model recovery error (MRE) [35], which is expressed as $\mathrm{MRE} = \sum_{i=1}^{n} \left| p(R_i) - p^{\mathrm{true}}(R_i) \right|$. The complete reproduction of $p^{\mathrm{true}}(R)$ yielded an MRE value of 0.

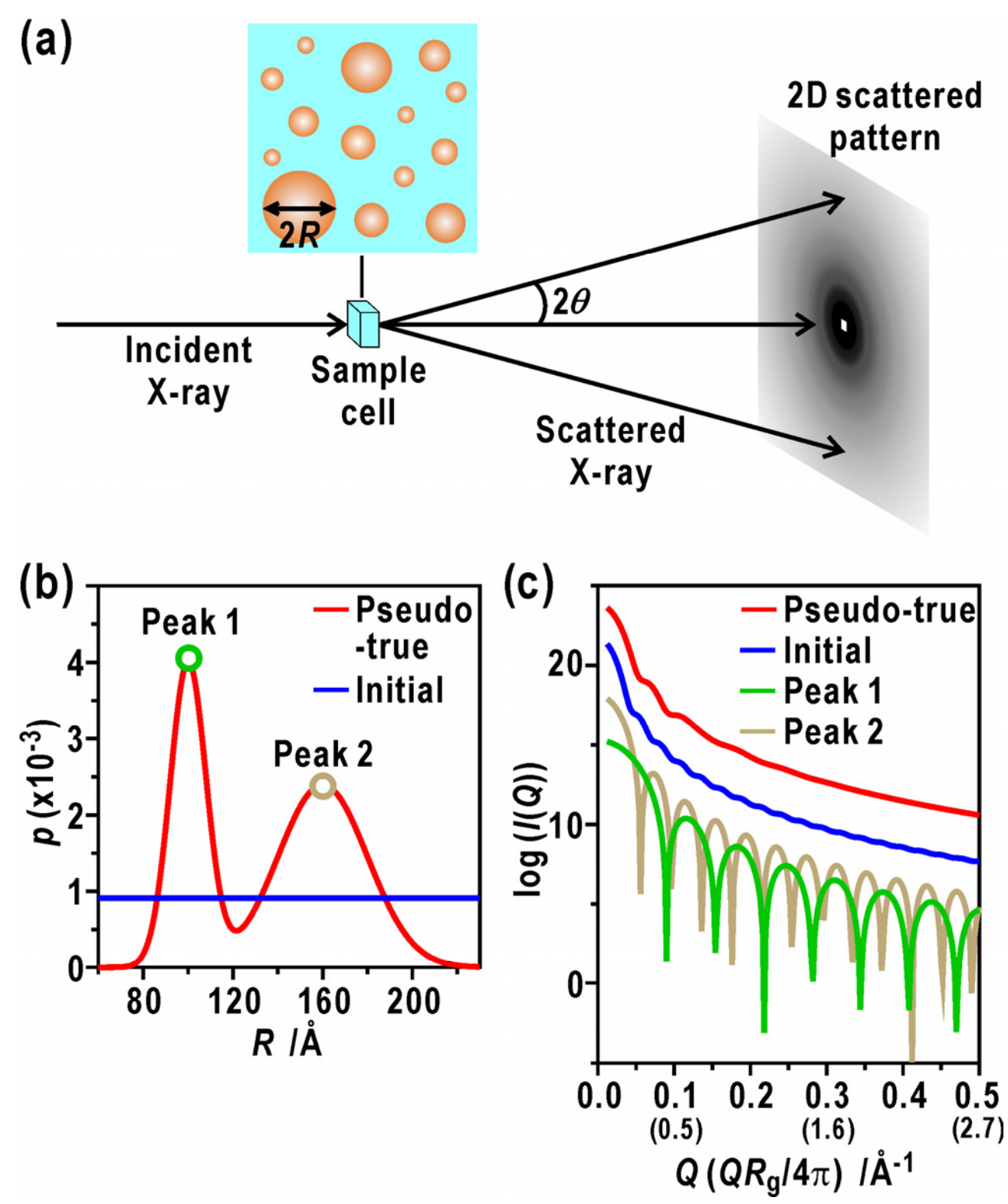

**FIG. 3.** Setup of the simulations that reconstruct particle size distribution (PSD) from pseudo-experimental SAXS data. (a) Schematic of the SAXS measurement for PSD in solution [28,29]. (b) Pseudo-true (red) and initial uniform (blue) PSDs. The particles at the peaks in the small and large $R$ areas are referred to as Peak 1 (green) and 2 (brown). (c) SAXS data for the pseudo-true PSD, initial PSD, Peaks 1 and 2.

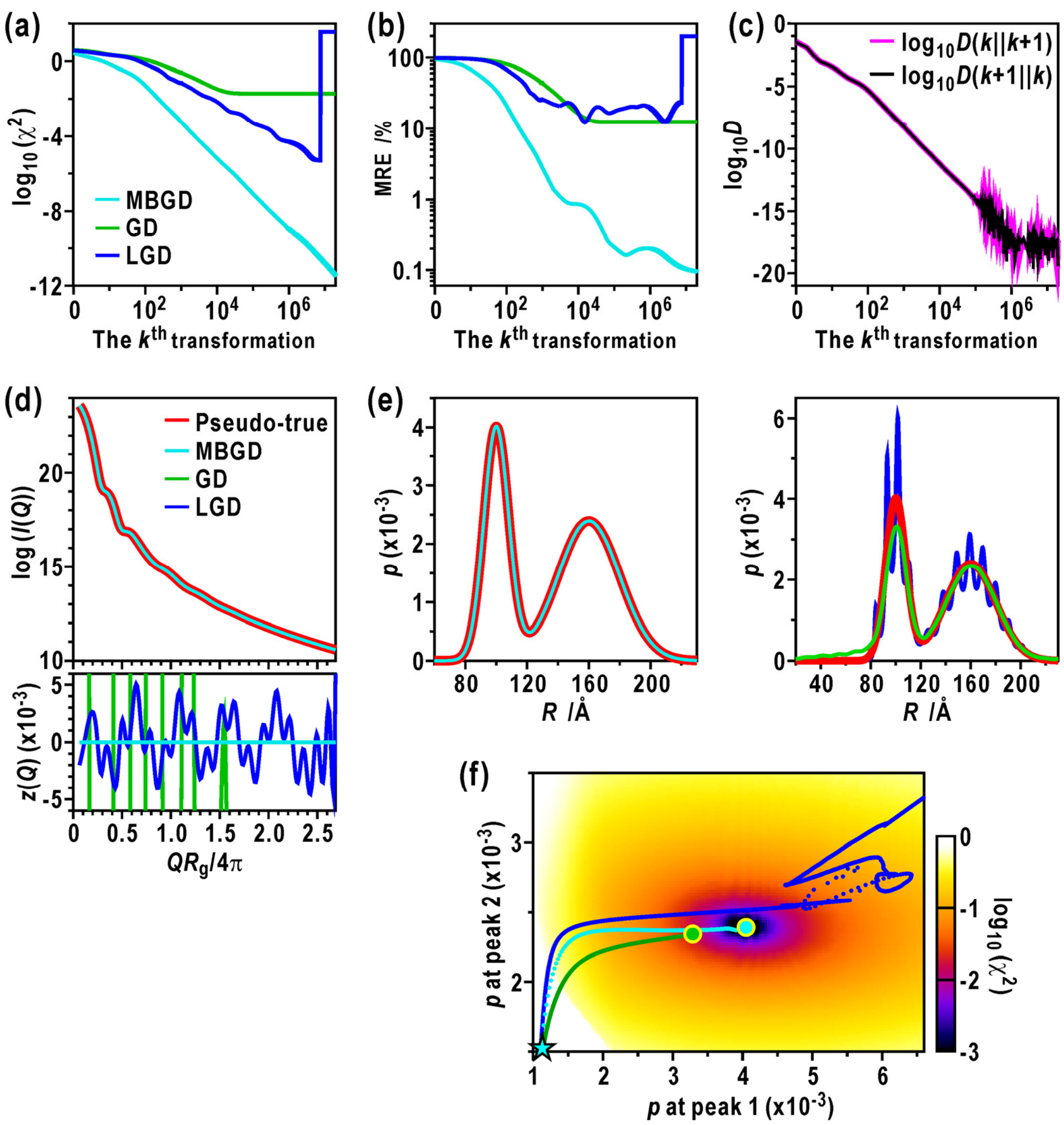

**FIG. 4.** Results of the simulations that reconstruct PSD using MBGD (cyan), GD (green), and LGD (blue). (a-b) The minimization processes monitored through (a) $\chi^2$ and (b) MRE. (c) Progress of forward (pink) and backward (black) KL divergences, $D(k||k+1)$ and $D(k+1||k)$, respectively, during the process by MBGD. (d) Comparisons of the pseudo-experimental SAXS data (red) and those obtained using MBGD, GD, and LGD. The residuals between the pseudo-experimental and calculated data were evaluated using $z(Q)=\left\{I^{\mathrm{CALC}}(Q)-I^{\mathrm{EXP}}(Q)\right\}/\sigma(Q)$ (lower). (e) Comparisons of the pseudo-true PSD and those reconstructed using MBGD (left), GD, and LGD (right). (f) Minimization processes projected on the landscape of the objective functional in a two-dimensional space, where the coordinates correspond to the PSD values at Peak 1 and Peak 2. The star and circle represent the initial and final points of the process in this space, respectively. The color represents the $\chi^2$ value (right bar).

In the reconstruction simulation of the PSD system, our MBGD method successfully minimized the objective function and accurately reconstructed the pseudo-true PSD, showcasing its theoretical validity (cyan curves in Fig. 4). The excellent ability of the MBGD to minimize the objective functional was validated by the fact that the $\chi^2$ value approached zero asymptotically (Figs. 4(a) and 4(d)). In terms of accurate reconstruction, the MRE converged to within 0.1% (Fig. 4(b)), showing nearly complete reproduction of the pseudo-true PSD by MBGD (Fig. 4(e)). From a practical perspective, MBGD is computationally efficient, requiring only a few seconds to reach 10,000 steps, during which the MRE falls below 1%. These results demonstrate the effectiveness of our manifold-based approach in the inverse problem of reconstructing PDs.

The evolution of the KL divergence during the minimization process also supports our theoretical finding that the KL divergence serves as a metric when the transformation of the PD is an infinitesimal displacement along an exponential geodesic. The equivalence between the forward and backward KL divergences $D(k||k+1)$ and $D(k+1||k)$, respectively, was maintained throughout this process (Fig. 4(c)). Furthermore, both of the KL divergences were equal to the second-order approximation $\sum_{i=1}^{n}\left(\Delta p_i^{\alpha}\right)^2 \big/ 2p_i^{\mathrm{O}}$ (Fig. S1(a) in [67]), confirming the validity of the approximation in Eq. (7). To intentionally violate this geometric condition, we increased the step size $\tau$, and beyond a certain threshold, the equivalence deteriorated and the minimization process failed (Fig. S1(b-d) in [67]). Furthermore, for values of $\tau$ below threshold, Eq. (7) remains valid. These outcomes further validate our findings regarding KL divergence. From a practical perspective, the upper limit for $\tau$ can be easily determined based on the aforementioned geological condition, Eq. (7), eliminating the need for an extensive search for $\tau$ (Note S1 in [67]).

To further assess the indispensability of our manifold-based approach, we performed reconstruction simulations using conventional gradient descent, assuming a Euclidean geometry in either $\mathbf{p}$ or log $\mathbf{p}$ spaces. These two formulations are expressed as follows:

$$\Delta\mathbf{p} \propto -\tau\nabla_{\mathbf{p}}\Phi\left(\mathbf{p}\right), \tag{18}$$

and

$$\Delta\log\mathbf{p} \propto -\tau\nabla_{\log\mathbf{p}}\Phi\left(\mathbf{p}\right). \tag{19}$$

Hereafter, we refer to Eqs. (18) and (19) as the GD and LGD methods, respectively. In these methods, we adopted the Lagrange multiplier method to satisfy the constraints imposed on **p**, as well as those of other studies [33,34]. Although both GD and LGD reduced the $\chi^2$ value (Fig. 4(a)), they failed to accurately reconstruct the pseudo-true PSD (Figs. 4(b) and 4(e)). Since the pseudo-experimental data here includes no noises, a correct method, such as MBGD, would be expected to achieve complete reproduction. Furthermore, even in their ability to minimize the $\chi^2$ values, both methods are inferior to MBGD. These behaviors of GD and LGD were independent of step size $\tau$.

To ascertain the reason for the failure of Euclidean-based approaches, such as GD and LGD, in achieving accurate reconstruction, we examined how these three methods differ in their minimization process on the landscape of the objective functional (Fig. 4(f)). To facilitate this analysis, we represented the landscape in a two-dimensional space, with the two coordinates corresponding to the PSD values at peaks 1 and 2 (Fig. 3(b)). The landscape was visualized by scanning the parameters of the mixture Gaussian model. As shown in Fig. 4(f), the minimization process in the MBGD successfully converged to the global minimum of the landscape. In contrast, the GD process became stuck before reaching the well containing the global minimum. In the LGD, the process initially approached the well but eventually diverged in an unintended direction. From a mathematical perspective, the gradient descent was formulated assuming that the objective function behaves like a potential function, generating a gradient field that directs the descent toward a minimum [68]. Therefore, the observed behaviors of the GD and LGD processes on the landscape demonstrate that by disregarding the manifold structure of the **p**-space, these methods compromise the potential function nature of the objective functional. In contrast, only the MBGD method, which properly considers the manifold structure, maintains this property. Consequently, the incorporation of manifold structure of p-space is essential for inverse problems of reconstructing PDs from experimental data.

### B. Evaluation of the ill-posedness of the inverse problem

As shown theoretically in Subsection II.C, the objective functional becomes convex when the experimental data are presented as expectation values of observables with respect to the PD, such as $I^{EXP}(Q)$. The convexity of the objective functional ensures a unique minimum, leading to results that are independent of the initial PD model (Fig. 2(b)). Indeed, in the reconstruction simulations, utilizing a sufficient amount of SAXS data up to $QR_g/4\pi = 2.69$ ($Q$ = 0.5 Å$^{-1}$), the MBGD calculations consistently converged to the true PSD regardless of the choice of the initial PSD models (Fig. 5(a)). This outcome validates the theoretical consideration regarding convexity. As discussed in Subsection II. C, the ill-posedness in the present inverse problem can be reframed as a concern about the sharpness of the minimum of the objective functional.

Subsequently, we explored how the sharpness of the minimum was affected by reducing the amount of information contained in the pseudo-experimental SAXS data (the amount of SAXS data). The results, demonstrating the independence of minimizations from the initial PSD, indicate that the objective functional is sufficiently sharp to yield a unique and well-defined minimum when using SAXS data up to $QR_g/4\pi = 2.69$. This implies that the SAXS data within this data range contain sufficient information for a unique PSD reconstruction. However, as the upper limit of $QR_g/4\pi$ was gradually decreased, thereby reducing the amount of SAXS data utilized, both the minimization and reconstruction performance deteriorated accordingly (Fig. 5(b) and 5(c), respectively). Then, the reconstruction accuracy deteriorated rapidly; the MRE exceeded 10%, when the upper limit fell below $QR_g/4\pi = 0.48$. Notably, this limit approximately corresponds to the position of the first dip in the scattering function of the particle corresponding to Peak 1 (Fig. 3(c)), indicating that at least the first oscillation must be observed as the spacing between the dips is inversely proportional to the particle size. These findings suggest that reducing the amount of information contained in the experimental data flattens the minimum of the objective, thereby increasing the ill-posedness of the problem. When utilizing SAXS data up to $QR_g/4\pi = 0.48$, the minimization process was stuck near the edge of the well containing the pseudo-true PSD, and the reconstruction results depended on the initial PSD models (Fig. 5(d)). Nevertheless, the reconstruction accuracy remained below 10%, which was still better than that achieved by GD or LGD using SAXS data up to $QR_g/4\pi = 2.69$.

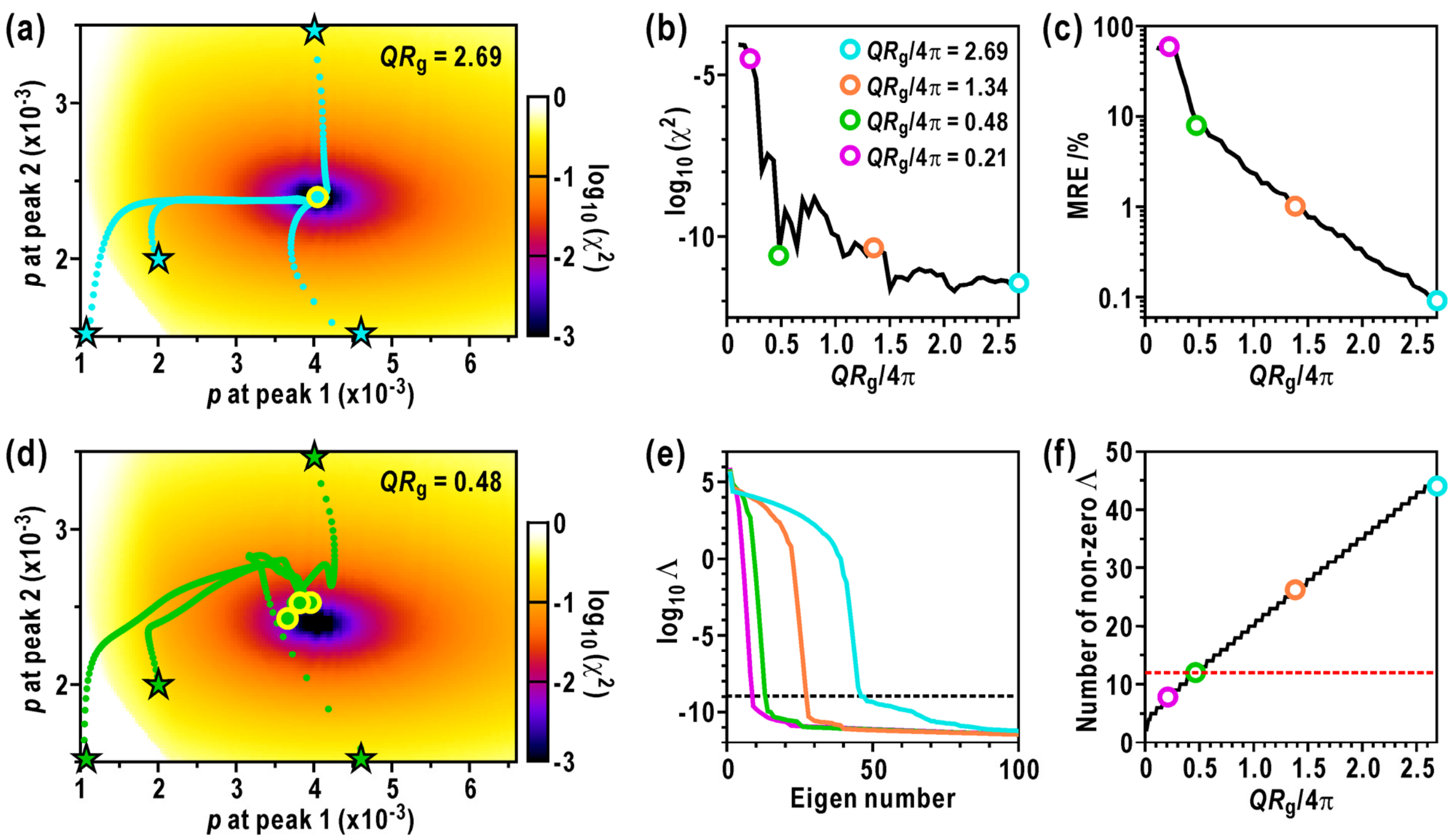

**FIG. 5.** Dependencies of the minimization and reconstruction accuracies on the amount of pseudo-experimental SAXS data. The amount of the data was reduced by decreasing the upper limit of $QR_g$. (a) Dependencies of the minimization processes on the initial PSD models when the SAXS data up to $QR_g = 2.69$ was utilized. These processes are projected on the landscape of the objective functional as well as in Fig. 4(f). The star and circle represent the initial and final points in these processes, respectively. (b) Dependencies of the minimized $\chi^2$ value on the amount of the pseudo-experimental SAXS data. (c) Dependencies of the reconstruction accuracy monitored by MRE on the amount of the data. (d) Dependencies of the minimization processes on initial PSD models, when the SAXS data up to $QR_g = 0.48$ was utilized. (e) Distributions of the eigenvalues of the Hessian of the objective functional. The upper limits of $QR_g$ are 2.69 (cyan), 1.34 (orange), 0.48 (green), and 0.21 (pink). The black dashed line represents the lower bound for the nonzero eigenvalues. (f) Dependencies of number of nonzero eigenvalues on the amount of data. The red dashed line represents the limit, below which the reconstruction accuracy becomes worse than 10%.

Next, to quantitatively evaluate the ill-posedness of the reconstruction, we examined the changes in the eigenvalue distribution of the Hessian with the reduction in the amount of SAXS data (Fig. 5(e)). A clear correlation is observed between the number of nonzero eigenvalues and ill-posedness (Figs. 5(c) and 5(f)). When using SAXS data up to $QR_g/4\pi = 2.69$, which provided a strictly sharp minimum of the objective functional, 44 nonzero eigenvalues were observed. This number decreased as the amount of SAXS data decreased, falling below 10 for $QR_g/4\pi \leq 0.38$. In this regime, the reconstruction failed with an accuracy worse than MRE = 20%. Consequently, the number of nonzero eigenvalues can serve as a valuable indicator for assessing the ill-posedness, even in inverse problems of PD reconstruction.

### C. Reconstruction simulation of protein conformational ensembles from SAXS data

Next, we conducted simulations in which protein conformational ensembles were reconstructed from the pseudo-experimental SAXS data. As a model system, we utilized transferrin comprising two domains [69,70] (Fig. 6(a)). Through AAMD and CGMD simulations (see Supplementary Methods S1 and S2, respectively, in [67]), we observed that the primary motion of transferrin is shown to be an open-close movement between the domains (domain motions) (Note S2 in [67]). In addition, the SAXS data of transferrin predominantly depended on the aforementioned open-close motion, allowing for the simplification the objective functional (Notes S3 and S4 in [67]). The red dashed line represents the lower bound for the nonzero eigenvalues of the ensemble as a one-dimensional PD based on the distance between the domains, $R_{\mathrm{CM}}$ (Fig. 6(a)), as well as the PSD system. To discretize a PD, $p(R_{\mathrm{CM}})$, we grouped conformations within each bin of 0.2 Å along $R_{\mathrm{CM}}$ into a single conformational state, as substantial variations in SAXS data were observed among states of this size (Note S4 in [67]). The resultant number of states became 65. Hereafter, we employed multiple-Gō CGMD simulations (Methods S2 in [67]) to generate model ensembles for the reconstruction simulations, for the following two reasons. First, since the structural information provided by SAXS data is of low resolution [71,72], the scattering profiles computed from CGMD ensembles are in good agreement with both experimental SAXS data and those calculated from AAMD simulations [73]. Second, CGMD is capable of generating multiple ensembles with distinct population distributions, making it particularly suitable for the present reconstruction simulations.

As a pseudo-true PD in reconstruction simulations, we generated a model PD with the primary populations located at open conformational states (Fig. 6(c)). Two types of PD models were prepared as initial PDs for the reconstruction simulations, (Fig. 6(c)): a uniform PD and a force field-based PD generated by CGMD. The uniform PD assumes a lack of prior knowledge regarding the physical principles governing protein motions, whereas the force field-based PD utilizes the same physical parameters as the pseudo-true PD but with different populations at closed conformational states. Despite the limitations of molecular force field accuracy, protein motions can be computationally reproduced to some extent using MD simulations [52]. The use of the force field-based PD as the initial ensemble virtually mimics this situation.

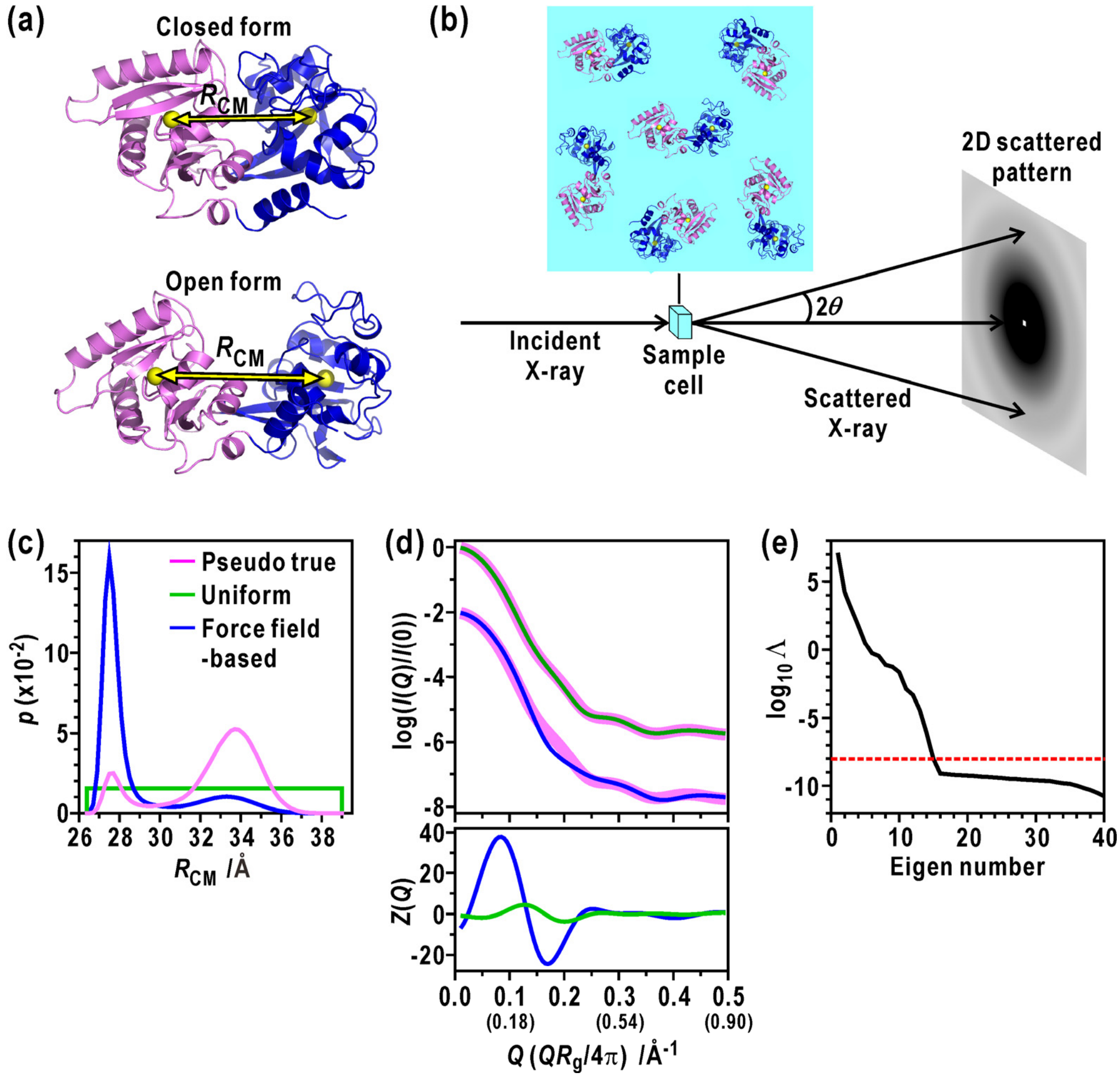

**FIG. 6.** Setup of the simulations that reconstruct protein conformational ensemble from pseudo-experimental SAXS data. (a) Transferrin as a model system. The crystal structures of the holo and apo forms are presented in the upper and lower panels, respectively (PDB IDs: 1A8E and 1BP5, respectively). The domain distance $R_{\mathrm{CM}}$ represents the distance between the mass centers of the two domains. (b) Schematic of the SAXS measurement for protein conformational ensembles in solution [71,72]. (c) Protein conformational ensembles can be represented as PDs on $R_{\mathrm{CM}}$ in the case of transferrin, showing the pseudo-true (pink), uniform (green), and force field-based (blue) PDs. (d) Comparisons of the SAXS data (upper panel) among these PD models. The residuals were evaluated using $z(Q)=\left\{I^{\mathrm{CALC}}(Q)-I^{\mathrm{EXP}}(Q)\right\}/\sigma(Q)$ (lower). (d) Distribution of the eigenvalues of the Hessian of the objective functional. The red dashed line represents the lower bound for the nonzero eigenvalues.

In the subsequent reconstruction simulations, the SAXS data calculated from the pseudo-true PD (Methods S3 in [67]) were utilized as the pseudo-experimental data, with the $\chi^2$ function serving as the objective functional. The experimental errors necessary for the calculation of $\chi^2$ were determined based on the analyses on the SAXS biological data bank [74] (Method S4 in [67]). Because experimental SAXS data $I^{EXP}(Q)$, represent the rotational and conformational averages of single-molecule X-ray scattering (Fig. 6(b)), they can be expressed as expected values [49,71,72]. This renders the objective functional convex. Significant differences in the SAXS data were observed among the pseudo-true, uniform, and force-field-based PDs (Fig. 6(d)), underscoring the effectiveness of SAXS in ensemble reconstruction. Notably, the number of nonzero eigenvalues of the Hessian of the $\chi^2$ function reached 14 when utilizing data up to $QR_g/4\pi = 0.90$ ($Q = 0.5$ Å$^{-1}$) (Fig. 6(e)). Based on the results of the reconstruction simulations using the PSD system (Fig. 5), while this number is deemed adequate for reconstructing the PD over $R_{CM}$, it may fall short of achieving complete reconstruction.

Consistent with the PSD system, the MBGD calculation applied to the conformational ensemble of transferrin succeeded in both reducing the $\chi^2$ value (Fig. 7(a) and 7(c)) and reconstructing the pseudo-true PD (Fig. 7(b) and 7(d)) when employing the pseudo-experimental SAXS data up to $QR_g/4\pi = 0.90$. However, the calculation results demonstrated a dependence on the initial PDs, likely stemming from the limited number of 14 non-zero eigenvalues. Initiating the MBGD calculation from the uniform PD yielded a reconstructed PD with an accuracy of MRE = 5%, which is considered satisfactory (upper panel in Fig. 7(d)); however, a complete reconstruction was not achieved. In contrast, initiating the calculation from the force field-based PD led to near-complete reconstruction with an accuracy surpassing 1% (the lower panel in Fig. 7(d)). In line with its superior reconstruction accuracy, the latter calculation yielded SAXS data that closely matched the pseudo- experimental data (Fig. 7(c)). We also conducted reconstruction simulations for other proteins exhibiting domain motions and obtained consistent results: superiority of the force-field-based PD as the initial input (Notes S5 and S6 in [67]). These results indicate that the current inverse problem of reconstructing the PDs of protein domain motion from SAXS data is inherently ill-posed. Nevertheless, utilizing a force field-based PD as the initial input can lead to near-complete reconstruction.

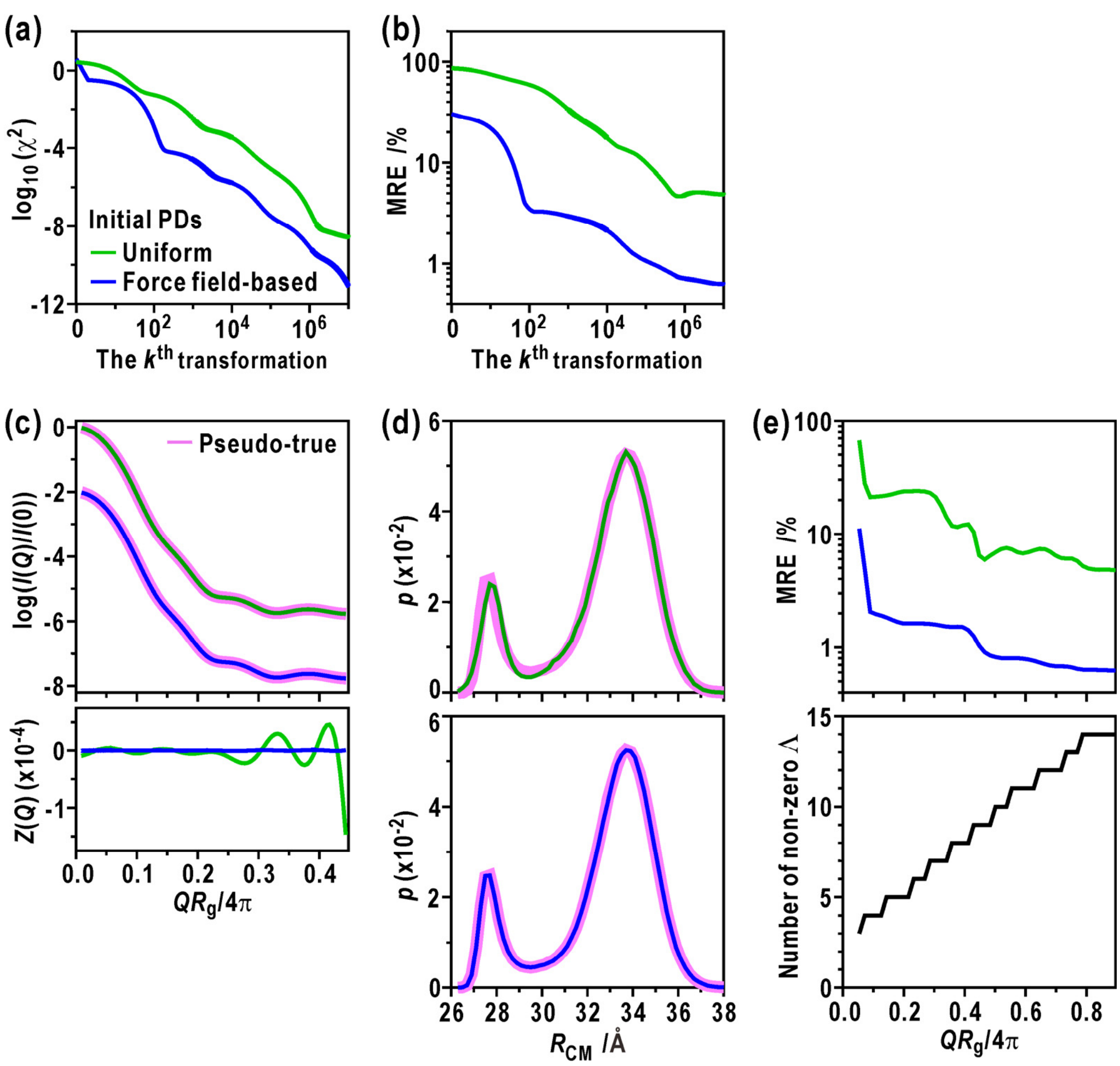

**FIG. 7.** Results of the simulations reconstructing conformational ensemble of transferrin from its pseudo-experimental SAXS data using the MBGD method. The green and blue represent the results for MBGD, which were initiated from the uniform and force field-based PDs. (a-b) Minimization processes monitored through $\chi^2$ (a) and MRE (b). (c) Comparisons of the pseudo-experimental SAXS data (red) and those obtained using MBGD. Residuals between the pseudo-experimental and calculated data evaluated using $z(Q)=\left\{I^{\mathrm{CALC}}(Q)-I^{\mathrm{EXP}}(Q)\right\}/\sigma(Q)$ (lower). (d) Comparisons of the pseudo-true PD and those reconstructed using MBGD. (e) Dependencies of the reconstruction accuracy monitored using MRE (upper) and the number of nonzero eigenvalues (lower) on the amount of data. The white-filled circles represent the upper bound of the data range. Narrowing the range further results in an MRE exceeding 10% (black dashed line).

Furthermore, we investigated the robustness of the reconstruction accuracy with respect to the flatness of the objective functional. Through reconstruction simulations with reduced amounts of pseudo-experimental SAXS data of transferrin (upper panel in Fig. 7(e)), we assessed the number of nonzero eigenvalues of the Hessian of the objective functional (lower panel in Fig. 7(e)) necessary to achieve reconstruction accuracy surpassing MRE = 10%. Remarkably, the force field-based PD model significantly improved the robustness of the reconstruction accuracy. When the uniform PD was used as the initial input, the required number of nonzero eigenvalues was nine. In contrast, when the force field-based PD was used, the required number was only four, which was significantly smaller than that required for the PSD system. However, the required number depends on the complexity of protein motion. In the case of protein domain motion with two degrees of freedom, initiating the MBGD calculation from the uniform PD failed in an accurate reconstruction of a pseudo-true PD even though the number of nonzero eigenvalues was 12 (Note S6 in [67]).

In this section, we explored the feasibility of applying the MBGD method to type (ii) systems, using protein conformational ensembles with SAXS data as a representative case. Despite the ill-posedness of the inverse problem when utilizing SAXS data up to $QR_g/4\pi = 0.90$, the results demonstrate that the MBGD method, when initialized with force field-based PDs, is highly effective in both minimizing the objective functional and accurately reconstructing the true PDs.

**D. Application of MBGD to experimental SAXS data of SjGlcNK**

Finally, we applied the MBGD method to actual experimental SAXS data to elucidate the conformational ensembles. The target sample was SjGlcNK, an enzyme that catalyzes the phosphorylation of D-glucosamine (GlcN) [65]. SjGlcNK is composed of the N-terminal cap, intermediate, and C-lobe domains (Fig. 8(a)). ATP binds to the active site located in the cleft between the intermediate and C-lobe domains, whereas GlcN binds to the C-lobe domain. X-ray crystallographic analyses revealed that the binding of ATP and GlcN induced domain-closure motion in SjGlcNK (upper panel in Fig. 8(b)), indicating significant conformational flexibility.

We conducted 2-μs AAMD simulations to explore the conformational space of SjGlcNK. Principal component analysis (PCA) [75] revealed that the first and second principal components (PCs) corresponded to the open-close and twisting motions of the C-lobe domain relative to the intermediate domain, respectively, in both the AAMD and CGMD trajectories (Fig. 8(b)). The twisting motion of the second PC was not observed in the crystallographic analyses [65]. The potential of the mean force (PMF) map indicated that the AAMD simulations explored a wider range of conformational space compared with that of the crystal structures (Fig. 8(d)).

We first conducted the MBGD calculation using the AAMD-derived ensemble as the initial input. However, this approach resulted in minimization failure and a reconstructed ensemble with density discontinuities at its edges, owing to insufficient conformational sampling by AAMD (Note S7 and Fig. S12 in [67]). Therefore, to broaden the sampling range, we conducted CGMD simulations using 36 representative structures extracted from the AAMD ensemble as reference conformational states (AA+CGMD) (Fig. 8(e)). The MBGD calculation utilizing this AA+CGMD-derived ensemble as the initial input successfully reproduced the experimental data (Fig. 8(c)) and reconstructed an ensemble with no discontinuities (Fig. 8(f)), indicating that the AA+CGMD simulations provided sufficiently broad sampling. Furthermore, the number of nonzero eigenvalues of the Hessian of the objective functional was nine. Based on the results of the reconstruction simulations presented in the previous section, this number suggests a reconstruction accuracy that surpasses MRE = 10%. It is noteworthy that the experimental SAXS data could not be explained without the twisting motion identified by AAMD. This result indicates that the realistic motions provided by AAMD are essential for accurate ensemble reconstruction.

The ensemble reconstructed by MBGD revealed that 71.2% of the population was distributed around the closed conformations observed in the crystal structure of the holo form, even in the absence of substrates. This indicates a population shift mechanism [76] for substrate recognition by different structures. (b) Domain motions of SjGlcNK identified using the principal component analysis (PCA) of the AAMD trajectory: the first PC mode (upper panel) and second PC mode (lower panel). The left and right panels represent the front and side views, respectively. The direction of $C_\alpha$ displacements along these PC modes are represented using blue arrows. (c) Comparison of the experimental (pink) and calculated SAXS data. The green and blue curves represent the data calculated from the AA+CGMD and reconstructed ensembles, respectively. (d-f) Ensembles projected on the first and second PC modes: the AAMD, AA+CGMD, and MBGD-reconstructed ensembles, respectively. The ensembles are visualized using a map plotting the potential of the mean force (PMF). The crystal structure of the holo form is represented by the pink circle, and those of the two apo forms are presented by the pink triangle and square. The structures utilized for the input of the AA+CGMD simulations are represented by the small pink diamonds. The two "dropped-jar" conformations are represented by the red circle and triangle. (g-j) Side views of the entire structure of the conformations shown in the PMF map (Fig. 8(f)): crystal structure of the holo form (pink circle in Fig. 8(f)), the crystal structure of the apo form (pink triangle), dropped-jar conformation 1 (red circle), and dropped-jar conformation 2 (red triangle), respectively. (k-n) Closer views of the active-site cleft in the aforementioned conformations. The order is consistent with that of the entire structure. The D118 residue is represented in purple.

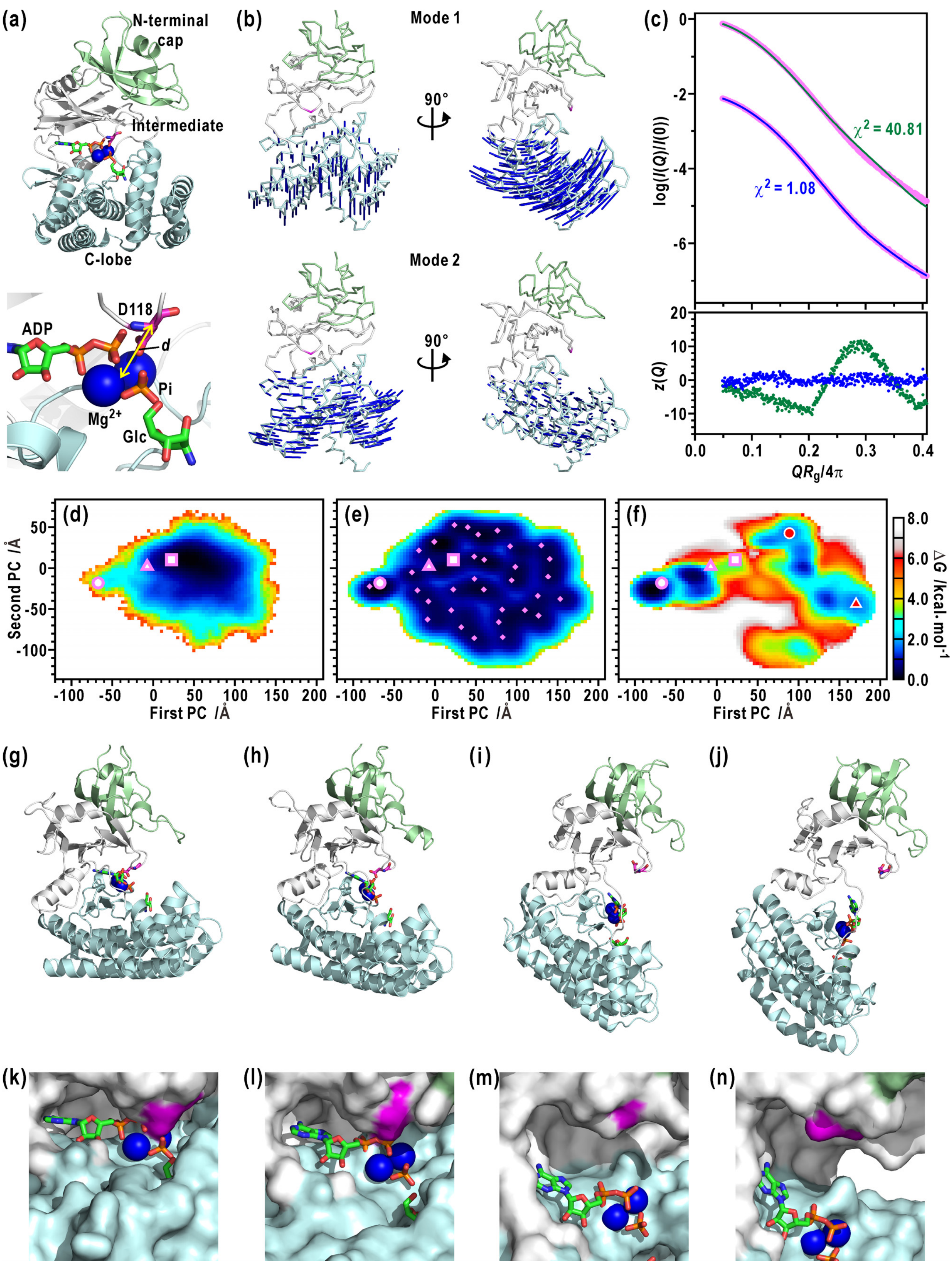

**FIG. 8.** Application of the MBGD method to experimental SAXS data of SjGlcNK under substrate-free conditions. (a) Crystal structure of the holo form of SjGlcNK (PDB ID: 6HWL); the entire structure (upper panel) and active-site cleft (lower panel). The domain structures are shown using

SjGlcNK. Furthermore, a substantial population was observed at conformations where the active-site cleft was more open than that in the crystal structures of the apo form, referred as a "dropped-jar" conformation hereafter (Figs. 8(i) and 8(j)). To validate the populations of these dropped-jar conformations, we conducted the MBGD calculation starting from an ensemble that excluded populations of these conformations. The results showed a reconstructed ensemble with discontinuities (Note S7 and Fig. S13 in [67]), demonstrating that the experimental data could not be considered without dropped-jar conformations. Notably, the closed and dropped-jar conformations were separated by a free-energy barrier not observed in the AAMD-derived ensemble, highlighting the need for improvements in the force field. Nevertheless, AAMD remains indispensable, as it can sample regions of the conformational space that are inaccessible by crystallographic analyses or CGMD alone.

The subsequent question concerns the reasons for the functional necessity of the dropped-jar conformations. A closer look at the conformational changes in the active-site cleft (Figs. 8(k-n)) can provide some clues. The cleft of the substrate-bound structure was so narrow that the catalytic residue D118 and two magnesium ions were sandwiched between the phosphate groups of the bound ATP (Fig. 8(k)). However, this sandwich-like structure was maintained in the half-closed crystal conformation of the apo form (Fig. 6(l)). Consequently, the binding of ATP to the clefts of these crystal structure conformations would likely be difficult owing to the prevention of ATP access by the sandwich-like structure. In contrast, the clefts in the dropped-jar conformations were sufficiently wide to allow ATP access (Figs. 8(i) and 8(j)). Based on these observations, we hypothesized that the dropped-jar conformations are crucial in facilitating ATP access to the active site of SjGlcNK.

In summary, the application of MBGD to the experimental SAXS data of SjGlcNK allows for the successful reconstruction of the conformational ensembles of this enzyme. The reconstruction results revealed a population-shift mechanism for substrate binding in this enzyme. Furthermore, through the reconstructed ensemble, we identified populations of ATP-accessible conformations that could not be detected through crystallographic analyses or MD simulations alone.

## IV. DISCUSSION

Observations of the GD and LGD processes in the landscape revealed that these standard gradient descent methods compromised the potential-function nature of the objective functional, unlike the MBGD method (Fig. 4(f)). To provide a theoretical explanation for this compromise, we consider the following setup. Two fixed points, O and A, exist in either the Euclidean space or a manifold, and the gradient field $\mathbf{F} = -\nabla \Phi$ of the objective functional $\Phi$ is integrated along two arbitrary paths, $C_1$ or $C_2$ (Fig. 9). In the case of the Euclidean space $\mathbf{x}$ (Fig. 9(a)), the potential-function property of $\Phi$ implies that the integral is independent of the integration path.

$$-\int_{C_1} \mathbf{F} \cdot d\mathbf{x} = -\int_{C_2} \mathbf{F} \cdot d\mathbf{x} = \Phi \qquad (20)$$

In contrast, on manifolds (Fig. 9(b)), owing to the curvature and torsion inherent in space [77], the integral of the gradient field $\mathbf{F}$ expressed in Eq. (20) depends on the choice of path and local coordinate systems, $\boldsymbol{\mu}$ and $\boldsymbol{\eta}$, such that

$$-\int_{C_1}\mathbf{F}\cdot d\boldsymbol{\mu} \neq -\int_{C_2}\mathbf{F}\cdot d\boldsymbol{\mu} \neq \varPhi \quad \text{or} \quad -\int_{C}\mathbf{F}\cdot d\boldsymbol{\mu} \neq -\int_{C}\mathbf{F}\cdot d\boldsymbol{\zeta} \neq \varPhi \tag{21}$$

Based on Eq. (21), neglecting the manifold structure of space compromises the potential-function nature of $\varPhi$. In the case of a general Riemannian manifold $\zeta$ with the metric tensor $\mathbf{G}(\zeta)$, the gradient can be calibrated as $\mathbf{G}^{-1}(\zeta)\nabla_{\zeta}\varPhi(\zeta)$ [78], allowing for the recovery of the equality in Eq. (20). However, implementing this correction in **p**-space remains challenging owing to the unknown normalization constant [79]. The MBGD method circumvents this challenge by incorporating the manifold structure with no gradient correction.

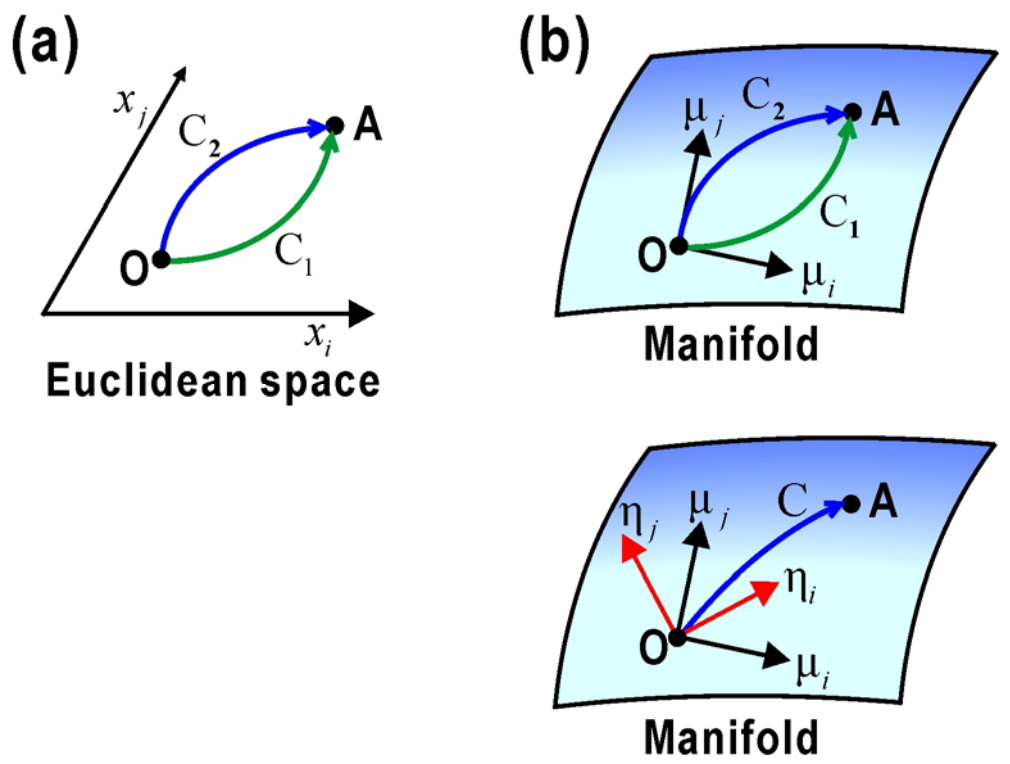

**FIG. 9.** Schematics of the line integration of gradient vector in Euclidean space (a) and manifold (b).

As an application to type (ii) systems, whose constituent elements undergo state transitions, we conducted the reconstruction of protein conformational ensembles from experimental SAXS data through the MBGD method. Our simulations have shown that accurate reconstruction with an error rate below 10% is achievable when a physics-based PD is used as the initial input to MBGD. These findings underscore the significance of incorporating *a priori* physical knowledge of system components in reconstructions. The effectiveness of physics-based PDs has also been demonstrated in other studies [33,34,54]. In the reconstruction of protein conformational ensembles, MD-derived ensembles have been used as initial models and have succeeded to some extent in explaining experimental data, even without the use of manifold-based techniques.

Furthermore, the reconstructed PD can contribute to refining the physical understanding of the system by comparing it with the physics-based PD model utilized as the initial input. In the case of SjGlcNK, the energy barrier between the open and closed conformational states observed in the reconstructed ensemble was not present in the AAMD-derived ensemble, meaning that the inter-domain interactions through water molecules between domains were not properly captured by the current molecular force field [80,81]. In future studies, accumulating such discrepancies between physics-based model and reconstructed ensembles will be beneficial for the development of more accurate molecular force fields. Nevertheless, AAMD is indispensable for obtaining physics-based conformational ensemble required for the reconstruction of the true ensembles as the initial input.

The analyses conducted in this study regarding the ill-posedness of inverse problems suggest that the number of nonzero eigenvalues of the Hessian of the objective functional can serve as a useful indicator of ill-posedness. Our findings indicate that approximately ten nonzero eigenvalues were necessary to achieve a reconstruction accuracy surpassing 10% in both the PSD system and ensembles of protein domain motions, although this may vary based on the initial PD model. Furthermore, we focused on the simple PD, which can be described with one or two degrees of freedom. However, the PDs in many physical systems are more complex, requiring higher degrees of freedom for the description. Such examples for biomolecular conformational ensembles include multi-domain proteins composed of more than two domains (MDPs) [82,83], intrinsically disordered proteins (IDPs) [84,85], and nuclear biomolecules [86,87]. In such complex systems, ten nonzero eigenvalues is expected to be insufficient to achieve an accurate reconstruction.

One effective strategy for addressing an insufficient number of eigenvalues is to increase the amount of experimental data (Fig. 2(b)). Based on the results of the reconstruction simulations, augmenting the amount of experimental data leads to an increase in the number of nonzero eigenvalues, thereby sharpening the minimum of an objective functional. Various experimental techniques, such as solution neutron scattering [50,88-89], nuclear magnetic resonance [90-93], double electron-electron resonance [35,94], and cryo-electron microscopy [95,96] are available for investigating protein conformational ensembles. Indeed, the integration of these diverse data sources yields significant benefits for the structural studies of IDPs [97-99]. Given these diverse data, we should assess the appropriate number of nonzero eigenvalues required for accurate reconstruction through simulations using synthetic data. Subsequently, manifold-based techniques such as MBGD could accurately y reconstruct PDs from these experimental data.

## V. CONCLUSIONS

In this study, we applied the information geometry framework to inverse problems for reconstructing PDs from experimental data. First, we theoretically demonstrated that the KL divergence between two PDs can serve as a distance when the transformation between them corresponds to an infinitesimal displacement along an e-geodesic on the **p**-space manifold. Based on these findings, we formulated the MBGD. The reconstruction simulations of the PSD system revealed that, unlike standard gradient descent methods, the MBGD method maintains the potential-function nature of the objective functional. This unique property enables MBGD to successfully achieve both the minimization of the objective functional and the accurate reconstruction of the true PD. Based on the simulation results, we applied the MBGD method to actual experimental SAXS data for SjGlcNK. This application successfully visualized a functionally important conformational ensemble that could not be obtained solely through crystallographic analyses, SAXS measurements, or MD simulations.

This study demonstrated that the inverse problem of reconstructing PDs from experimental data is fundamentally solvable when the manifold structure of **p**-space is properly considered. PDs in type (ii) systems, particularly in soft matter [30-35], pose challenges in direct observation, leading to limited data accumulation. Consequently, the application of machine learning to such systems remains challenging, making inverse problem approaches, such as those presented in this paper, essential for visualizing PDs. The collection of data acquired through these methods is anticipated to enhance our comprehension of physical systems and facilitate the application of machine learning techniques. We believe that manifold-based techniques, such as MBGD, will be crucial in expanding the applicability of inverse problem approaches for visualizing PDs in type (ii) systems beyond protein conformational ensembles.

**Acknowledgements**

This study was supported by grants from the Japan Society for the Promotion of Science (26104535, 26800227, 17K19209, 17H04854, and 21K03489 to T. O.; 18H05229 to T. O., R. I., and M. S.).

**Data availability**

The code and the data used in the reconstruction simulations are available on GitHub [100].

# APPENDIX

## Appendix A: Taylor expansion of the KL divergence

The Taylor expansions of the forward and backward KL divergences from point O to point α in **p**-space, with respect to the displacement $\Delta\mathbf{p}^{\alpha}$, can be written as follows:

$$D\left(\mathrm{O}\,\|\,\alpha\right)=\sum_{i=1}^{n}\Delta p_{i}^{\alpha}+\sum_{i=1}^{n}\sum_{j=2}^{\infty}\frac{\left(-1\right)^{j}}{j}\cdot\frac{\left(\Delta p_{i}^{\alpha}\right)^{j}}{\left(p_{i}^{\mathrm{O}}\right)^{j-1}} \tag{A1}$$

and

$$D\left(\alpha\,\|\,\mathrm{O}\right)=\sum_{i=1}^{n}\Delta p_{i}^{\alpha}+\sum_{i=1}^{n}\sum_{j=2}^{\infty}\frac{\left(-1\right)^{j}}{\left(j-1\right)j}\cdot\frac{\left(\Delta p_{i}^{\alpha}\right)^{j}}{\left(p_{i}^{\mathrm{O}}\right)^{j-1}}, \tag{A2}$$

respectively. Therefore, assuming $\Delta\mathbf{p}^{\alpha}$ to be infinitesimal, the second-order approximation of the above expressions leads to Eq. (7).

## Appendix B: Proof of Pythagorean Theorem for points on e- and m-geodesics

Here, we demonstrate that the points on e- and m-geodesics satisfy the Pythagorean Theorem, as indicated in Eq. (9) when these geodesics are orthogonal. The proof utilizes points, O, α, and β, as referenced in the main text. By substituting Eqs. (4), (6), and (8) into the Fisher information metric, Eq. (2), we obtain

$$\sum_{i=1}^{n-1}\left(\theta_{i}^{\alpha}-\theta_{i}^{\mathrm{O}}\right)\left(\eta_{i}^{\beta}-\eta_{i}^{\mathrm{O}}\right)=0\,. \tag{B1}$$

By substituting Eq. (3), the left-hand side of Eq. (A1) can be expressed using KL divergence as follows:

$$
\begin{aligned}
\sum_{i=1}^{n-1}\left(\theta_i^{\alpha}-\theta_i^{\mathrm{O}}\right)\left(\eta_i^{\beta}-\eta_i^{\mathrm{O}}\right) &= \left(\sum_{i=1}^{n-1}\theta_i^{\alpha}\eta_i^{\beta}+\log p_n^{\alpha}\right)-\left(\sum_{i=1}^{n-1}\theta_i^{\alpha}\eta_i^{\mathrm{O}}+\log p_n^{\alpha}\right) \\
&\quad -\left(\sum_{i=1}^{n-1}\theta_i^{\mathrm{O}}\eta_i^{\beta}+\log p_n^{\mathrm{O}}\right)+\left(\sum_{i=1}^{n-1}\theta_i^{\mathrm{O}}\eta_i^{\mathrm{O}}+\log p_n^{\mathrm{O}}\right) \\
&= \left\{\sum_{i=1}^{n-1}p_i^{\beta}\log p_i^{\alpha}+\left(1-\sum_{i=1}^{n-1}p_i^{\beta}\right)\log p_n^{\alpha}\right\}-\left\{\sum_{i=1}^{n-1}p_i^{\mathrm{O}}\log p_i^{\alpha}+\left(1-\sum_{i=1}^{n-1}p_i^{\mathrm{O}}\right)\log p_n^{\alpha}\right\} \\
&\quad -\left\{\sum_{i=1}^{n-1}p_i^{\beta}\log p_i^{\mathrm{O}}+\left(1-\sum_{i=1}^{n-1}p_i^{\beta}\right)\log p_n^{\mathrm{O}}\right\}+\left\{\sum_{i=1}^{n-1}p_i^{\mathrm{O}}\log p_i^{\mathrm{O}}+\left(1-\sum_{i=1}^{n-1}p_i^{\mathrm{O}}\right)\log p_n^{\mathrm{O}}\right\}. \\
&= \sum_{i=1}^{n}p_i^{\beta}\log p_i^{\alpha}-\sum_{i=1}^{n}p_i^{\mathrm{O}}\log p_i^{\alpha}-\sum_{i=1}^{n}p_i^{\beta}\log p_i^{\mathrm{O}}+\sum_{i=1}^{n}p_i^{\mathrm{O}}\log p_i^{\mathrm{O}} \\
&= -\sum_{i=1}^{n}p_i^{\beta}\log\frac{p_i^{\beta}}{p_i^{\alpha}}+\sum_{i=1}^{n}p_i^{\mathrm{O}}\log\frac{p_i^{\mathrm{O}}}{p_i^{\alpha}}+\sum_{i=1}^{n}p_i^{\beta}\log\frac{p_i^{\beta}}{p_i^{\mathrm{O}}} \\
&= -D\left(\beta\,\|\,\alpha\right)+D\left(\mathrm{O}\,\|\,\alpha\right)+D\left(\beta\,\|\,\mathrm{O}\right)
\end{aligned}
$$

(B2)

Therefore, the Pythagorean Theorem in Eq. (9) is obtained by substituting Eq. (B2) and Eq. (7) into Eq. (B1).

## Appendix C: Geometrics of gradient descent path in general Euclidean space

This section delves into the geometric properties of the gradient descent path in general Euclidean space $\xi$. To accomplish this, we analyze the geometric relationship between two points $\xi^k$ and $\xi^{k+1}$ on the path, representing the positions after the $k^{\mathrm{th}}$ and $k+1^{\mathrm{th}}$ transformations through gradient descent, respectively (Fig. 10). The gradient descent formulation can be expressed using the Euclidean distance between these points, $d(k||k+1)$, as follows:

$$\nabla_{\xi^k}\left(\Phi\left(\xi^k\right)-d\left(k+1\,\|\,k\right)/\tau\right)=0\,. \tag{C1}$$

Mathematically, Eq. (B1) is equivalent to determining the extremum of a Lagrangian of the form:

$$L=\Phi\left(\xi^k\right)+\left(a^{k+1}-d\left(k+1\,\|\,k\right)\right)/\tau \tag{C2}$$

where $a^{k+1}$ represents a parameter to be determined. In the general Lagrange multiplier method, $\xi^{k+1}$

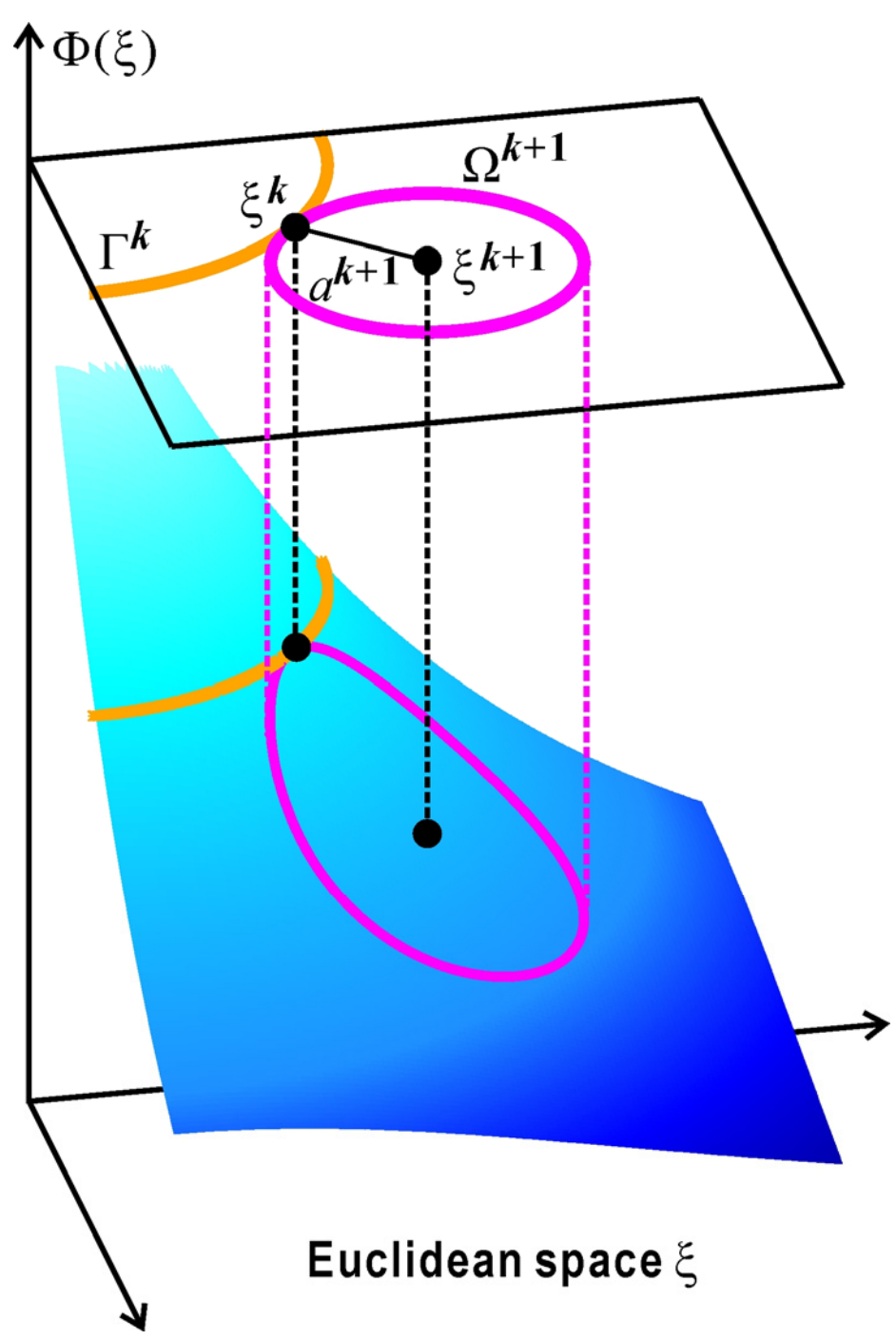

**FIG. 10.** Geometry of gradient descent path in general Euclidean space ξ.

and $a^{k+1}$ are known, whereas $\xi^k$ and $1/\tau$ denote the position and multiplier to be determined, respectively. In this scenario, by determining the extremum of the Lagrangian, Eq. (C2), is equivalent to determining the extremum of $\Phi(\xi)$ on the hypersphere $\Omega^{k+1}$ defined by the constraint $a^{k+1} - d\left(k+1\|\,k\right) = 0$ (Fig. 10). However, in the gradient descent, the relationship between the variables is reversed: $\xi^k$ and $1/\tau$ are known, whereas $\xi^{k+1}$ and $a^{k+1}$ represent the position and parameter to be determined, respectively. Therefore, when $\Gamma^k$ is defined as the contour of $\Phi(\xi)$ that passes through $\xi^k$, the task of determining the extremum of the Lagrangian, Eq. (C2), is equivalent to identifying the hypersphere $\Omega^{k+1}$ that is tangential to $\Gamma^k$ at point $\xi^k$ (Fig. 10). This represents the geometric relationship that must be maintained between the two points $\xi^k$ and $\xi^{k+1}$ along the gradient descent path. We assume that this geometric relationship holds true for the two points $\eta^k$ and $\eta^{k+1}$ along the gradient path in the **p**-space, particularly when these points demonstrate an approximately Euclidean relationship. This assumption is expressed using Eq. (12) in the main text.

# Supplementary Material for "Manifold-based transformation of the probability distribution: application to the inverse problem reconstructing distributions from experimental data"

Tomotaka Oroguchi[a,b*], Rintaro Inoue[c], and Masaaki Sugiyama[c]

[a]Department of Physics, Faculty of Science and Technology, Keio University, 3-14-1 Hiyoshi, Kohoku-ku, Yokohama 223-8522, Japan.
[b]RIKEN SPring-8 Center, 1-1-1 Kouto, Sayo-cho, Sayo-gun, Hyogo 679-5148, Japan
[c]Institute for Integrated Radiation and Nuclear Science, Kyoto University, Kumatori, Sennan-gun, Osaka 590-0494 Japan.

*Corresponding author
oroguchi@phys.keio.ac.jp

## SUPPLEMENTARY METHODS

### Method S1: All-atom MD simulations

The all-atom MD (AAMD) simulation was conducted using the AMBER16 [1] software package with the AMBER ff15FB [2] force field and TIP3PFB [3] water model. The crystal structures of proteins were utilized as the initial structures. The solution was prepared by placing the crystal structure in a truncated octahedron box and adding a 30 Å layer of water molecules. $Na^+$ and $Cl^-$ ions were introduced to neutralize the net charge of the system. Electrostatic interactions were handled using the particle-mesh Ewald method [4] with a real-space cutoff of 10 Å. Lennard–Jones interactions were truncated beyond 10 Å using a continuum model correction. Bonds involving hydrogen atoms were constrained using the SHAKE method [5], and a time step of 2 fs was set. Initially, the system was subjected to energy minimization of 1,000 steps. Subsequently, the temperature of the system was gradually increased from 10 K to 293 K in an NPT run of 300 ps at 1 atm. Finally, a 1-μs NVT run was conducted for each production run. Simulations were conducted using NVIDIA TITAN V GPUs.

### Method S2: Coarse-grained MD simulations

For CGMD simulations, we used the off-lattice Gō model employed in the CafeMol software [6]. In this model, ach amino acid residue was represented as a single CG particle located at the position of the $C_\alpha$ atom of the residue. The potential function is expressed as follows:

$$
\begin{aligned}
V_{\text{Gō}}\left\langle \mathbf{x} \middle| \mathbf{x}_0 \right\rangle = & \sum_{i=1}^{n-1} K_{\text{bond}}\left(b_{i,i+1} - b_{i,i+1,0}\right) + \sum_{i=1}^{n-2} K_{\text{angle}}\left(\theta_{i,i+1,i+2} - \theta_{i,i+1,i+2,0}\right) \\
& + \sum_{i=1}^{n-3} \left\{ \begin{aligned} & K_{\text{dihedral}}^{(1)}\left\{1 - \cos\left(\phi_{i,i+1,i+2,i+3} - \phi_{i,i+1,i+2,i+3,0}\right)\right\} \\ & + K_{\text{dihedral}}^{(3)}\left\{1 - \cos 3\left(\phi_{i,i+1,i+2,i+3} - \phi_{i,i+1,i+2,i+3,0}\right)\right\} \end{aligned} \right\} \\
& + \sum_{i<j-3}^{\text{native contact}} \varepsilon_{\text{Gō}} \left\{ 5\left(\frac{r_{ij,0}}{r_{ij}}\right)^{12} - 6\left(\frac{r_{ij,0}}{r_{ij}}\right)^{10} \right\} \\
& + \sum_{i<j-3}^{\text{non-native}} \varepsilon_{\text{ev}} \left(\frac{r_{ij,\text{repulsive}}}{r_{ij}}\right)^{12}
\end{aligned} \tag{S1}
$$

where $\mathbf{x}_0$ represents the 3D coordinates of the input structure. The bonded parameters are as follows: $b_{i,i+1}$ and $b_{i,i+1,0}$ represent the bond length between two neighboring residues and the corresponding length in $\mathbf{x}_0$, respectively; $\theta_{i,i+1,i+2}$ and $\theta_{i,i+1,i+2,0}$ denote the angle between three consecutive residues and the corresponding angle in $\mathbf{r}_0$, respectively; $\phi_{i,i+1,i+2,i+3}$ and $\phi_{i,i+1,i+2,i+3,0}$ denote the dihedral angle between four consecutive residues and corresponding dihedral angle in $\mathbf{x}_0$, respectively, with spring constants set to the values utilized in CafeMol [6]. The fourth term denotes the Gō potential that maintains the conformational topology of $\mathbf{x}_0$. The Gō term was calculated for any residue pair within a distance of $r_{\text{native}}$ in $\mathbf{x}_0$ (native contact pair). $r_{ij}$ denotes the distance between native contact pairs $i$ and $j$, whereas $r_{ij,0}$ denotes the distance in $\mathbf{x}_0$. The cutoff distance $r_{\text{native}}$ and force constant $\varepsilon_{\text{Gō}}$ were adjusted to maintain the rigidity of a domain structure (root mean squared deviations of CG particles ≤ 1.5 Å) while enabling structural flexibility, such as domain motions during simulations. The fifth term represents the repulsive potential between residue pairs $i$ and $j$, calculated for non-native contact pairs. The reference distance $r_{ij,\mathrm{repulsive}}$ depended on the amino acid residue types of $i$ and $j$, and the cutoff distance $r_{\mathrm{repulsive}}$ was set to 2.0. The force constant $\varepsilon_{\mathrm{ev}}$ was set to 0.5 kcal/mol/Å$^{12}$.

To generate a conformational ensemble that spanned apo and holo forms in each protein system, we utilized a multiple-Gō CGMD simulation, with the potential expressed as follows:

$$V_{\text{multi-Gō}}\left\langle \mathbf{x} \middle| \mathbf{x}_1, \mathbf{x}_2, \cdots, \mathbf{x}_L \right\rangle = -k_{\mathrm{B}}T \cdot \log\left\{ \sum_{i=1}^{L} \exp\left( -\frac{V_{\text{Gō}}\left\langle \mathbf{x} \middle| \mathbf{x}_i \right\rangle + v_i}{k_{\mathrm{B}}T_i} \right) \right\}, \tag{S2}$$

where $L$ is the number of input conformations and $\mathbf{x}_i$ is the input coordinate of the $i^{\mathrm{th}}$ conformation. $k_{\mathrm{B}}$ and $T$ are the Boltzmann constant and the simulation system temperature, respectively. $T$ was set to 293 K in all multiple-Gō CGMD simulations in this study. Parameter $v_i$ determines the relative stability of the $i^{\mathrm{th}}$ conformation against the other conformations. The height of the free-energy barrier between the $i^{\mathrm{th}}$ and $j^{\mathrm{th}}$ conformations is determined by parameters $T_i$ and $T_j$.

The ensembles of transferrin for reconstruction simulations were created using multiple-Gō CGMD simulations. The crystal structures of the apo [7] (PDB ID: 1BP5) and holo [8] (PDB ID: 1A8E) forms were utilized as inputs to represent stable open and closed conformational states, respectively.

**Method S3: Calculation of SAXS data from CGMD simulations**

We utilized the CGMD-SAXS method, as outlined in our previous study [9], to calculate the SAXS data from individual CG structures within the CGMD simulations. This method considered X-ray scattering from both the solvent excluded by a protein molecule (solvent-excluded volume) and hydration shell, which is crucial for accurately calculating X-ray scattering from protein solutions. The previous study [9] demonstrated that the CGMD-SAXS data were consistent with not only experimental data but also the SAXS data calculated from AAMD trajectories.

To accurately calculate X-ray scattering from the solvent-excluded volume, the solvent parameter required is the electron density of the solvent in the bulk region ($\rho$). For the hydration shell, the necessary parameters are the thickness and solvent density (hydration density) of the hydration layer. In the CGMD-SAXS method, the space surrounding a protein CG structure was divided into voxels with a side length $\Delta D$ of 3 Å. Subsequently, voxels within the hydration layer were selected based on their distance from the CG structure. The scattering factor from each voxel of the hydration layer is expressed as follows:

$$f_{\mathrm{hydration}}^{\mathrm{CG}}\left(Q\right) = w\Delta D^3 f_{\mathrm{water}}^{\mathrm{CG}}\left(Q\right) \tag{S3}$$

where $w$ represents the increase in the number density of water molecules in the hydration layer from the bulk density. $f_{\mathrm{water}}^{\mathrm{CG}}\left(Q\right)$ denotes the scattering factor of a single water molecule in an all-atom representation (one oxygen and two hydrogen atoms) and is expressed as follows:

$$f_{\text{water}}^{\text{CG}}\left(Q\right)=\left[\sum_{i,j=1}^{3} f_i\left(Q\right) f_j\left(Q\right)\frac{\sin\left(Qr_{ij}\right)}{Qr_{ij}}\right]^{1/2} . \tag{S4}$$

To calculate Eq. (S4), we utilized the coordinates of the TIP3P water model were used. In the present reconstruction simulations, the solvent density parameters $\rho$ and $w$ were set to 0.338 e/Å$^3$ and $0.112\times10^{-2}$ molecules/Å$^3$, respectively. Those values provide the good agreement with experimental SAXS data for model proteins [9]. All CGMD-SAXS calculations were conducted using NVIDIA TITAN V GPUs.

**Method S4: *Q*-dependent experimental errors of SAXS data**

To investigate the $Q$-dependence of experimental errors, which is necessary for the calculation of $\chi^2$ (Eq. (16)), we initially analyzed experimental SAXS data stored in the small angle scattering biological data bank [10]. All analyzed data were collected using modern single-photon counting detectors and varied in mass units and protein concentrations. Across all experimental data analyzed, the $Q$-dependence of the experimental errors $\sigma^{\text{EXP}}(Q)$ can be approximated using the following model function:

$$\sigma^{\text{model}}\left(Q\right)=aI_{\text{Guinier}}^{\text{EXP}}\left(0\right)\left\{I\left(Q\right)/I_{\text{Guinier}}^{\text{EXP}}\left(0\right)\right\}^{b} , \tag{S5}$$

where $I_{\text{Guinier}}^{\text{EXP}}\left(0\right)$ denotes the forward-scattering intensity estimated from the Guinier plot [11]. The used parameters for the pseudo-experimental SAXS data in the reconstruction simulations of protein conformational ensembles were $a$ = 0.001 and $b$ = 0.3, which are typical values observed in the analyzed experimental data.

## SUPPLEMENTARY NOTES

### Note S1: Dependence of MBGD results on step size $\tau$

Here, we investigated how the MBGD reconstruction depends on the step size $\tau$, using the PSD system in Section A of the main text. For all values of $\tau$ up to a certain threshold (the red dash line in Fig. S1(b-d)), the MBGD method successfully minimized the objective functional and accurately reconstructed the pseudo-true PSD (Fig. S1(b) and (c)). For these values of $\tau$, the equivalence between the forward and backward KL divergences was also maintained (Fig. S1(d)). However, beyond the threshold, the equivalence deteriorated and the minimization process failed. These results demonstrated our findings that the approximate metric nature of the KL divergence is necessary to maintain the potential function nature of the objective functional.

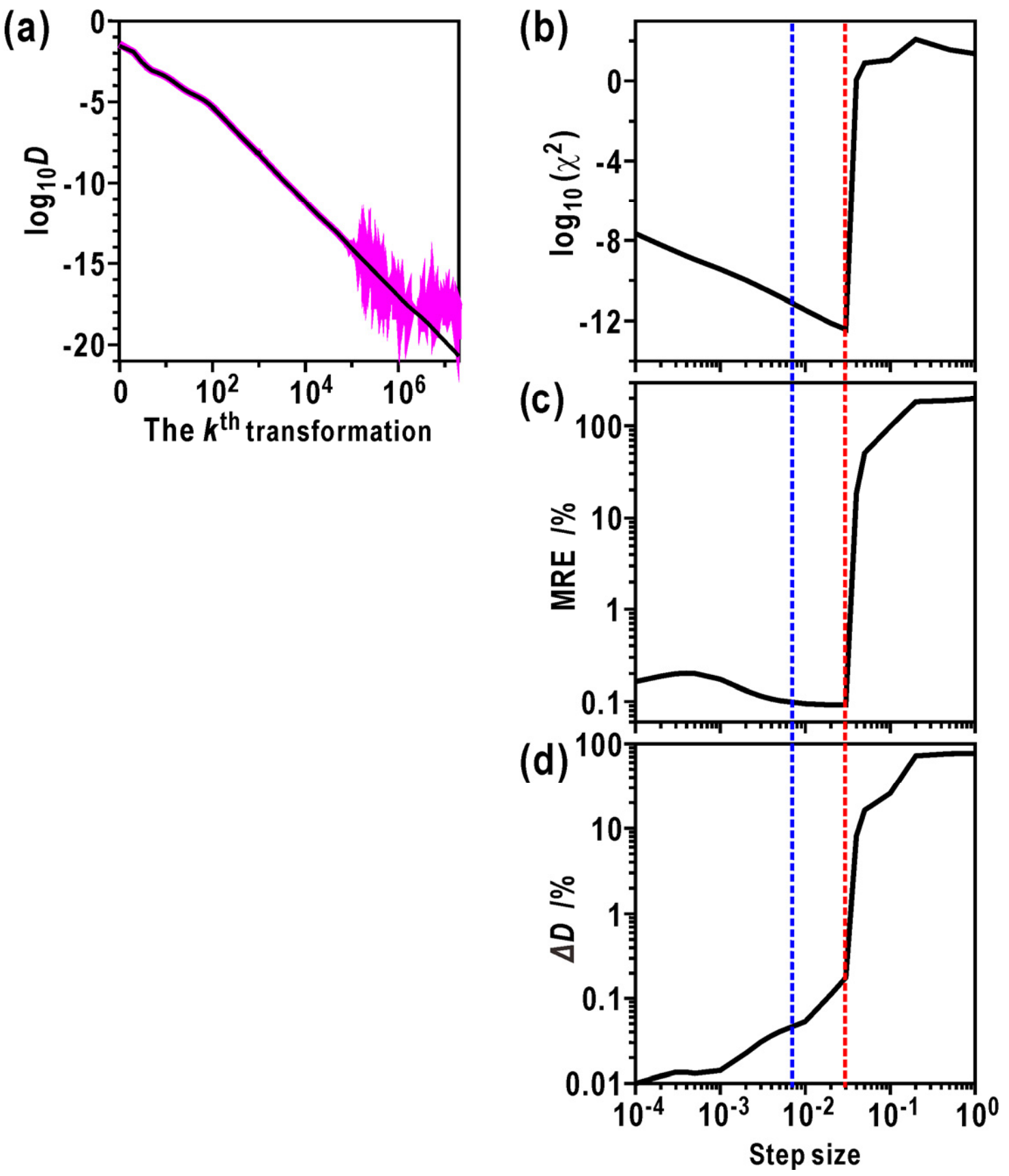

**FIG. S1.** Dependencies of the simulation results reconstructing PSD on the step size $\tau$. (a) Progress of forward KL divergence (pink) and the second-order approximation $\sum_{i=1}^{n}\left(\Delta p_i^{\alpha}\right)^2 \Big/ 2p_i^{\mathrm{O}}$ (black). (b) Dependencies of the minimized $\chi^2$ value on $\tau$. (c) Dependencies of the reconstruction accuracy monitored by AREP on $\tau$. (d) Dependencies of the equivalence between the forward and backward KL divergences on $\tau$. The equivalence was evaluated by $\Delta D = \left|D\left(k \,\|\, k+1\right) - D\left(k+1 \,\|\, k\right)\right| \Big/ D\left(k \,\|\, k+1\right)$.

From a practical perspective, it is convenient if the above upper limit for $\tau$ can be estimated from the geological condition for the metric nature of the KL divergence. As such a condition, we considered the criterion that the third order term in the Taylor expansions of the backward KL divergence (Eq. (A2)) can be neglected compared to the second order term. This condition is expressed as follows:

$$\sum_{i=1}^{n} \frac{\left(\Delta p_i^k\right)^3}{6\left(p_i^k\right)^2} \Bigg/ \sum_{i=1}^{n} \frac{\left(\Delta p_i^k\right)^2}{2 p_i^k} \le \varepsilon \tag{S6}$$

where $p_i^k$ is the probability density for the $i^{\text{th}}$ state after the $k^{\text{th}}$ transformation, and $\Delta p_i^k = p_i^{k+1} - p_i^k$.

From the MBGD equation, Eq. (1), $\Delta p_i^k \big/ p_i^k$ can be expressed as follows:

$$\frac{\Delta p_i^k}{p_i^k} = \exp\left(-\tau \frac{\partial \Phi\left(\mathbf{p}^k\right)}{\partial p_i^k}\right) \Bigg/ \sum_{i=1}^{n} p_i^k \exp\left(-\tau \frac{\partial \Phi\left(\mathbf{p}^k\right)}{\partial p_i^k}\right) - 1 . \tag{S7}$$

By expanding the above expression in a Taylor series up to the second order in $\tau$, we obtain the followings:

$$\frac{\Delta p_i^k}{p_i^k} \simeq \tau\left(A_i^k + B_i^k \tau\right), \tag{S8}$$

where

$$A_i^k = \sum_{j=1}^{n} p_j^k \frac{\partial \Phi}{\partial p_j^k} - \frac{\partial \Phi}{\partial p_i^k} \tag{S9}$$

and

$$B_i^k = \frac{1}{2}\left(\frac{\partial \Phi}{\partial p_i^k}\right)^2 - \frac{\partial \Phi}{\partial p_i^k} \sum_{j=1}^{n} p_j^k \frac{\partial \Phi}{\partial p_j^k} + \left(\sum_{j=1}^{n} p_j^k \frac{\partial \Phi}{\partial p_j^k}\right)^2 - \sum_{j=1}^{n} p_j^k \left(\frac{\partial \Phi}{\partial p_j^k}\right)^2 . \tag{S10}$$

By substituting Eqs. (S8-S10), Eq. (S6) becomes

$$\tau \sum_{i=1}^{n} p_i^k \left(A_i^k + B_i^k \tau\right)^3 \Big/ 3 \sum_{i=1}^{n} p_i^k \left(A_i^k + B_i^k \tau\right)^2 \le \varepsilon . \tag{S11}$$

The second-order approximation of the Taylor series of the left side of (S12) becomes

$$D^k \tau^2 + E^k \tau - F^k \le 0 , \tag{S12}$$

where

$$D^k = 3 \sum_{i=1}^{n} p_i^k \left(A_i^k\right)^2 B_i^k - 3\varepsilon \sum_{i=1}^{n} p_i^k \left(B_i^k\right)^2 , \tag{S13}$$

$$E^k = \sum_{i=1}^{n} p_i^k \left(A_i^k\right)^3 - 6\varepsilon \sum_{i=1}^{n} p_i^k A_i^k B_i^k , \tag{S14}$$

and

$$F^k = 3\varepsilon \sum_{i=1}^{n} p_i^k \left(A_i^k\right)^2 \tag{S15}$$

Since Eq. (S15) ensures that $F^k$ remains positive, an upper bound for $\tau$ can be estimated from the following expression:

$$\tau \le \frac{-E^k + \left\{ \left(E^k\right)^2 + 4D^k F^k \right\}^{1/2}}{2D^k}. \qquad \text{(S16)}$$

The blue dash line in Fig. S1(b-d) represents the upper bound, which is estimated from Eq. (S16) using the initial uniform PD as $\mathbf{p}^k$. Despite being based on a second-order approximation in $\tau$, the estimation provides a value close to the actual upper bound.

**Note S2: Analyses on conformational motions of transferrin**

Upon binding of an iron ion, transferrin undergoes the domain motion from an open to a closed conformations [7,8], suggesting that this motion of transferrin also occurred in the solution. To characterize the conformational motions of transferrin in the AAMD and CGMD simulation, we performed principal component analysis [12] (PCA) on the trajectories of these simulations. In these PCAs, the motions of the C-domain against the N-domain were evaluated using the $C_\alpha$ best fits for the N-domain.

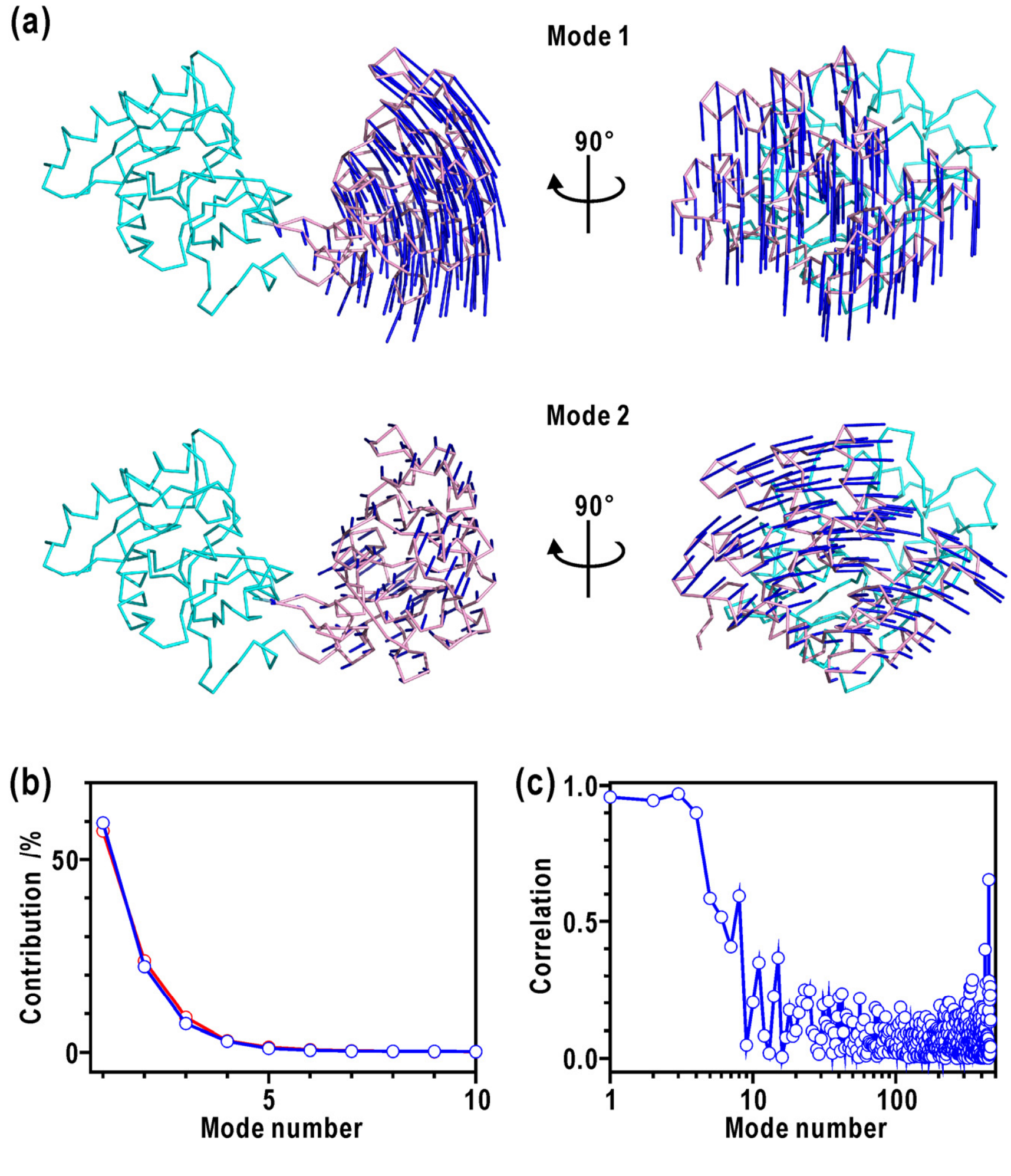

**FIG. S2.** Comparison of PCA in the AAMD and CGMD simulations of transferrin. (a) PCA modes calculated from AAMD trajectories: first PC mode (upper) and second PC mode (lower). The amplitude ($3\sigma$) and direction of $C_\alpha$ displacements along the first PC mode are represented by blue arrows. The domain structures are represented using different colors. (b) Individual contributions of the first 10 PC modes to total $C^{\alpha}$-mean square fluctuation in the AAMD (red) and CGMD (blue) simulations. (c) Correlation of the individual PC modes between AAMD and CGMD simulations.

The first and second PC modes accounted for approximately 80% of the total $C_\alpha$ mean square fluctuations in both the AAMD and CGMD simulations (Fig. S2(b)). These modes represented open-close and twisting motions of the C-domain against the N-domain, respectively (Fig. S2(a)). Furthermore, strong correlations were observed for the first four PC modes between the AAMD and CGMD simulations (correlation coefficient > 0.9), and certain correlations persisted up to the $8^{\text{th}}$ PC mode (Fig. S2(c)). These results indicate that the domain motions between the AAMD and CGMD simulations were consistent, validating the effectiveness of the CGMD simulation in capturing large amplitude motions.

**Note S3: Determination of projection coordinates and state resolution for representing probability distribution**

In inverse problems, the choice of coordinates describing a system, on which PDs are projected, must be determined based on the observables given as experimental data. Specifically, if the observables depend on only a subset of coordinates, it is meaningless to attempt to reconstruct the PD on other, irrelevant coordinates. For example, in the case of proteins, if NMR chemical shift data reflecting rotamer states of amino-acid residues are used as observables, it would be essentially impossible to reconstruct a PD defined over large-scale conformational coordinates, such as domain motions. Conversely, if SAXS data reflecting the global shape of the molecule are used as observables, reconstructing the PD over side-chain torsional coordinates would be infeasible. Therefore, when utilizing an inverse problem approach, it is essential to first identify which parameters of the system are effectively probed by the observables. Then, a meaningful coordinate system for PD reconstruction can be chosen.

After selecting the coordinate system, it is necessary to determine the optimal size of each state along that coordinate. A key requirement for a state was that all conformations within it yielded consistent values of observables. Although smaller state sizes satisfied this requirement, excessively small sizes unnecessarily increased the number of states, which was not ideal from a computational cost standpoint. Therefore, the size suitable for the reconstruction calculations was the largest among those that satisfied the aforementioned condition.

An alternative approach is the method which samples structures are treated as distinct states, and the corresponding weights are fitted individually [13]. However, under this approach, many degenerate structures—that is, conformations that yield indistinguishable values of the selected observables—are treated as separate states. This unnecessarily increases the number of states *n*, leading to reduced computational efficiency. Based on the rationale that structures indistinguishable in terms of the observables should be grouped into the same state, we chose not to adopt this approach.

**Note S4: Dependencies of the calculated SAXS data of transferrin on conformational changes**

To identify suitable coordinates for reconstructing transferrin conformational motions from SAXS data, we explored the correlation between the calculated SAXS data and conformational changes of this protein. The PC modes obtained through the PCA performed on the CGMD simulation (Fig. S3) were utilized to examine the conformational changes. Among these PC modes, the calculated SAXS data demonstrated dependencies only for the first two PC modes, which corresponded to domain motions. Therefore, we analyzed the dependence of the data on the 2D plane formed by these modes (Fig. S3(a)). All conformations within a 5 Å grid on this plane were categorized as a single conformational state represented by this grid. The SAXS data for each conformational state were calculated as the average of the data for all conformations within that state. This calculation utilized all conformations in the two ensembles generated by the multiple-Gō CGMD simulations.

To assess the dependencies of the SAXS data, we conducted an analysis by calculating the $\chi^2$ values for all pairs of conformational states. This process involved selecting a single conformational state on a plane, generating pseudo-experimental data from that state with Gaussian noise using $a$ = 0.001 and $b$ = 0.3 (Eq. (S5)), and then calculating $\chi^2$ between the pseudo-experimental data and noiseless SAXS data from other conformational states of the pair (Fig. S3(b)). In the calculation, the unknown experimental parameters related to solvent density and intensity scaling were fixed. An example of the $\chi^2$ plot on the 2D plane, which was calculated using the reference conformational state indicated by the red circle as the source of pseudo-experimental data is shown in Fig. S3(a). This plot indicated that the SAXS data depended on conformational changes along the first and second PC modes. However, a distinct region with consistently low $\chi^2$ values, forming a banded shape, was also observed (Figs. S3(a)). Within this banded-shaped region, the distance between the centers of mass of the two domains, $R_{\mathrm{CM}}$, remained relatively constant (Fig. S3(c)). These characteristics were observed for any conformational state within the 2D plane, indicating that the SAXS data for transferrin primarily depended on $R_{\mathrm{CM}}$. Therefore, $R_{\mathrm{CM}}$ was utilized as the coordinate to describe the conformational motions of transferrin.

Furthermore, we delved into determining the optimal state size for the coordinate $R_{\mathrm{CM}}$ in the reconstruction simulations. We divided the coordinate $R_{\mathrm{CM}}$ into sufficiently small subsets of 0.05 Å and computed the $\chi^2$ values for all subset pairs using the aforementioned procedure (Fig. S3(d)). The $\chi^2$ values between the subsets separated by > 0.2 Å exceeded 1.5; therefore, we selected 0.2 Å as the state size. The resultant number of states became 65.

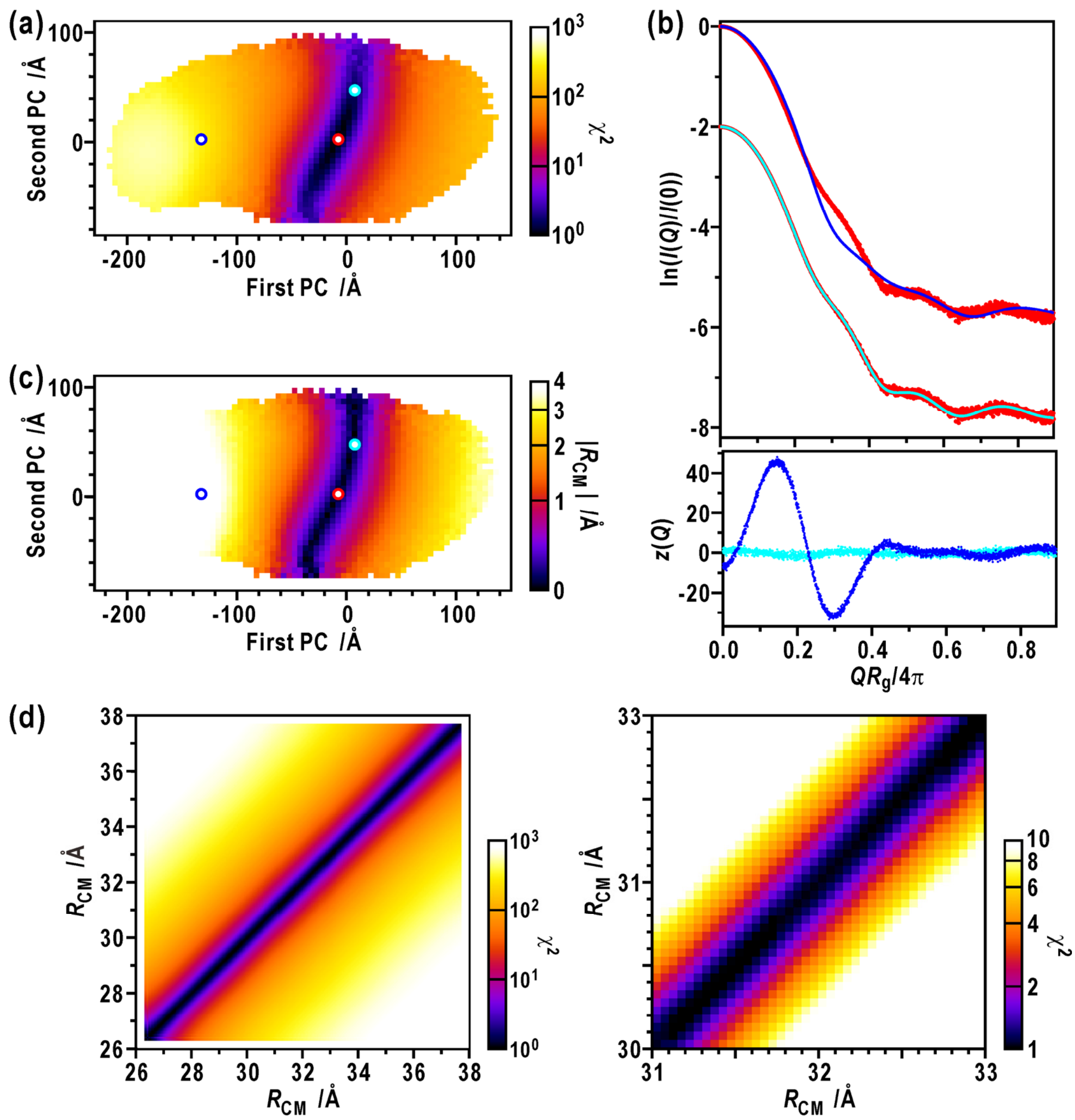

**FIG. S3.** Dependencies of the calculated SAXS data on conformational changes of transferrin. (a) Dependencies of $\chi^2$ values on domain motions outlined by the first and second PC modes. The red circle indicates the conformation from which the pseudo-experimental SAXS data were generated. (b) Comparison of the pseudo-experimental SAXS data (red) with the data calculated from the two representative conformations, representing those indicated by the blue and cyan circles in (a). (c) Dependencies of the coordinate $R_{CM}$ on the domain motions outlined by the first and second PC modes. The absolute value of the difference in $R_{CM}$, $|\Delta R_{CM}|$, from the conformation utilized to generate pseudo-experimental SAXS data (red circle) was plotted by color. (d) $\chi^2$ values between all pairs of conformations on $R_{CM}$, indicating the entire range (left panel) and magnified range (right panel) of $R_{CM}$.

### Note S5. Reconstruction simulation of conformational ensembles of *Mt*EPSPS from SAXS data

As the second model system for the validation test of the MBGD method, we employed *Mt*EPSPS [20]. The *Mt*EPSPS structure comprises two domains (Fig. S4(a)), and the crystal structures revealed that substrate binding induces domain-closure motion. We conducted the CGMD simulation using the crystal structure of the apo form (PDB ID: 2BJB [21]) for the Gō model potential. To characterize the conformational motions in the CGMD simulation, we analyzed the trajectory through PCA. Most of the total $C^{\alpha}$-mean square fluctuations were accounted for by the first and second PC modes (54 and 24%, respectively). These modes represented domain motions, which were open-close and twisting motions of the two domains, respectively (Fig. S4(a)).

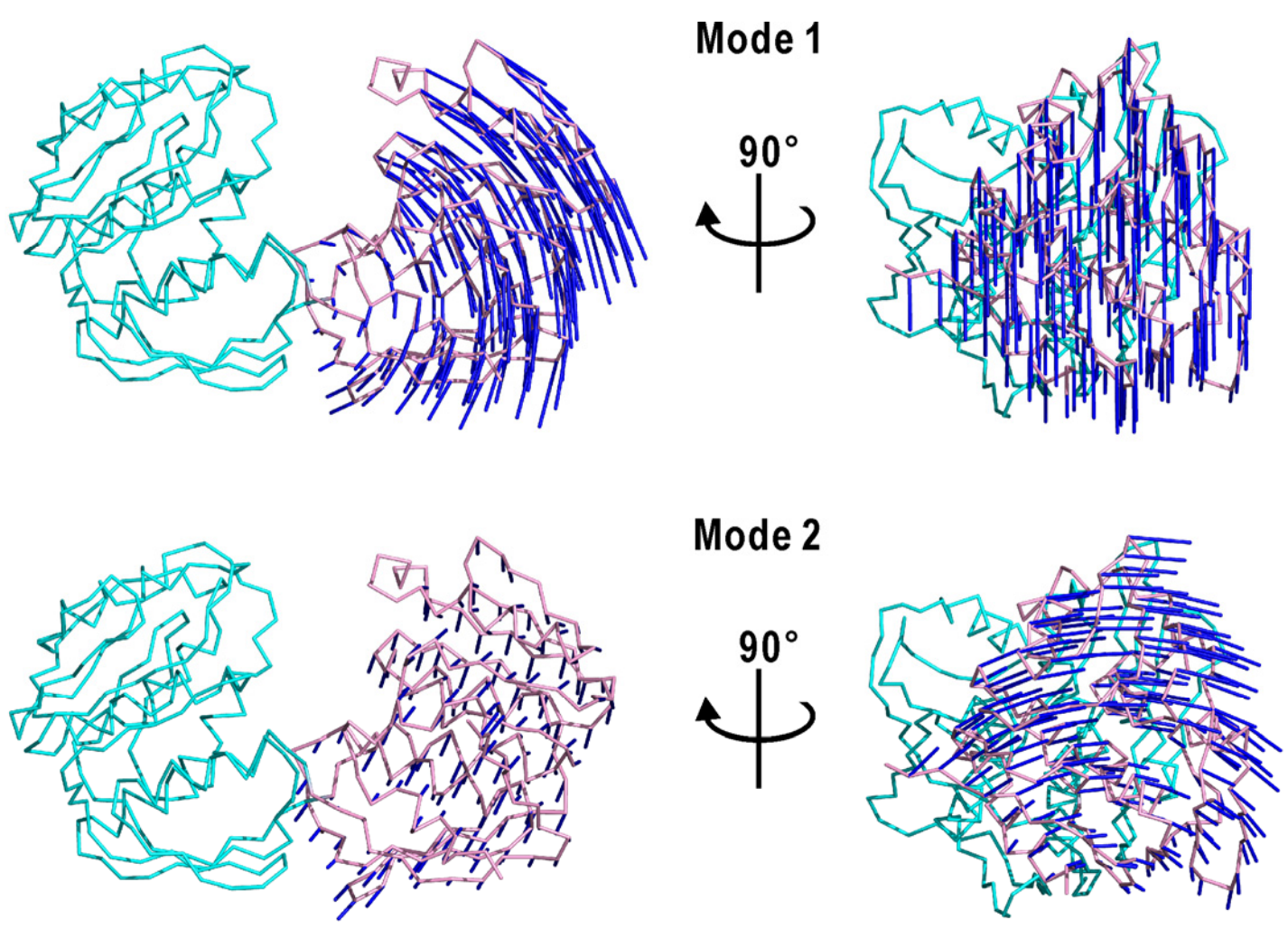

**FIG. S4.** Conformational motions of *Mt*EPSPS. PCA modes calculated from CGMD trajectories: first PC mode (upper) and second PC mode (lower). The amplitude ($3\sigma$) and direction of $C_{\alpha}$ displacements along the first PC mode are represented by blue arrows. The domain structures are represented using different colors.

The model PDs of *Mt*EPSPS for the reconstruction simulations were generated through the multiple-Gō CGMD simulation (Supplementary Methods). The crystal structures in the holo form (PDB ID: 2O0D [22]) and the apo form served as inputs for the metastable open and closed conformational states, respectively. Adjusting parameters in the multiple-Gō model potential produced two broad PDs, primarily populated in closed and open conformational states (Fig. S4(a)). These ensembles were used as the force field-based and pseudo-true PDs, respectively.

To select the coordinates on which PDs are projected, we investigated the dependence of the calculated SAXS data on the conformational changes of MtEPSPS. To describe the conformational changes, we used the PC modes obtained through the PCA conducted on the CGMD simulation mentioned previously. SAXS data dependencies were observed for the first two PC modes, representing domain motions. Therefore, we observed the dependence of the data on the 2D plane spanned by the first and second PC modes. All conformations within a 5 Å grid on the plane were categorized as a single conformational state represented by this grid. To observe the dependencies of the SAXS data, we calculated the $\chi^2$ values for all pairs of conformational states following the same procedure used for transferrin (Note S4).

Fig. S5(a) shows an example of the $\chi^2$ plot on the 2D plane, which was calculated using the reference conformational state indicated by the red circle as the state providing the pseudo-experimental data. As shown in this plot, the SAXS data depended on conformational changes along both the first and second PC modes. As well as the case of transferrin, there is a specific region where the $\chi^2$ value from the reference conformational state remains small (Fig. S5(a) and S5(b)), and it has a banded shape as observed for transferrin. However, in contrast to transferrin, the distance between the centers of mass of the two domains, $R_{\mathrm{CM}}$, did not remain the same within this banded region (Fig. S5(c)). Instead, we used the coordinate $R_{\mathrm{DM}}$, which was calculated as the average distance between all pairs of CG particles belonging to different domains. The $R_{\mathrm{DM}}$ value remained almost the same within the banded region (Fig. S5(d)). These characteristics of the $\chi^2$ values between the conformational state selected as a reference and the other states were observed for any conformational state in the 2D plane. These results demonstrate that the SAXS data of *Mt*EPSPS mainly depend on $R_{\mathrm{DM}}$. Therefore, we used cc as the coordinate to describe the conformational motions of *Mt*EPSPS. When utilizing $R_{DM}$ and SAXS data up to $QR_g/4\pi = 0.96$ ($Q = 0.5$ Å$^{-1}$), the number of nonzero eigenvalues of the Hessian of the $\chi^2$ function was 15 (Fig. S5(e)).

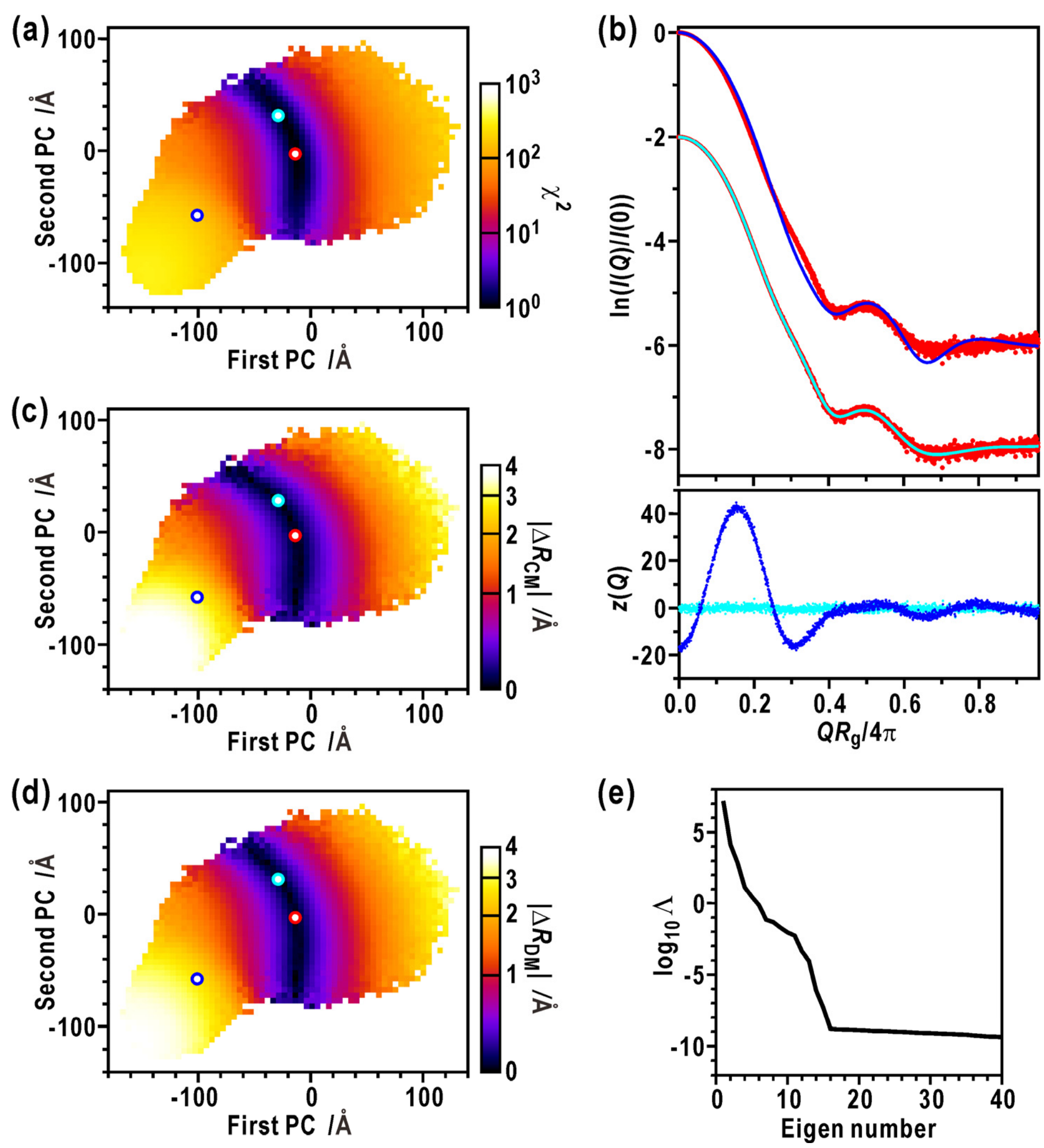

**FIG. S5.** Dependencies of the calculated SAXS data on conformational changes of *Mt*EPSPS. (a) Dependencies of $\chi^2$ values on domain motions outlined by the first and second PC modes. The red circle indicates the conformation from which the pseudo-experimental SAXS data were generated. (b) Comparison of the pseudo-experimental SAXS data (red) with the data calculated from the two representative conformations, representing those indicated by the blue and cyan circles in (a). Residuals between the data were evaluated using $z(Q)$ (lower). (c) Dependencies of the coordinate $R_{\mathrm{CM}}$ on the domain motions outlined by the first and second PC modes. The absolute value of the difference in $R_{\mathrm{CM}}$, $|\Delta R_{\mathrm{CM}}|$, from the conformation utilized to generate pseudo-experimental SAXS data (red circle) was plotted by color. (d) Dependencies of the coordinate $R_{\mathrm{DM}}$ on the domain motions outlined by the first and second PC modes. The absolute value of the difference in $R_{\mathrm{DM}}$, $|\Delta R_{\mathrm{DM}}|$, from the conformation utilized to generate pseudo-experimental SAXS data (red circle) was plotted by color. (f) Distribution of the eigenvalues of the Hessian of the objective functional when utilizing $R_{\mathrm{DM}}$ to describe the conformational motions.

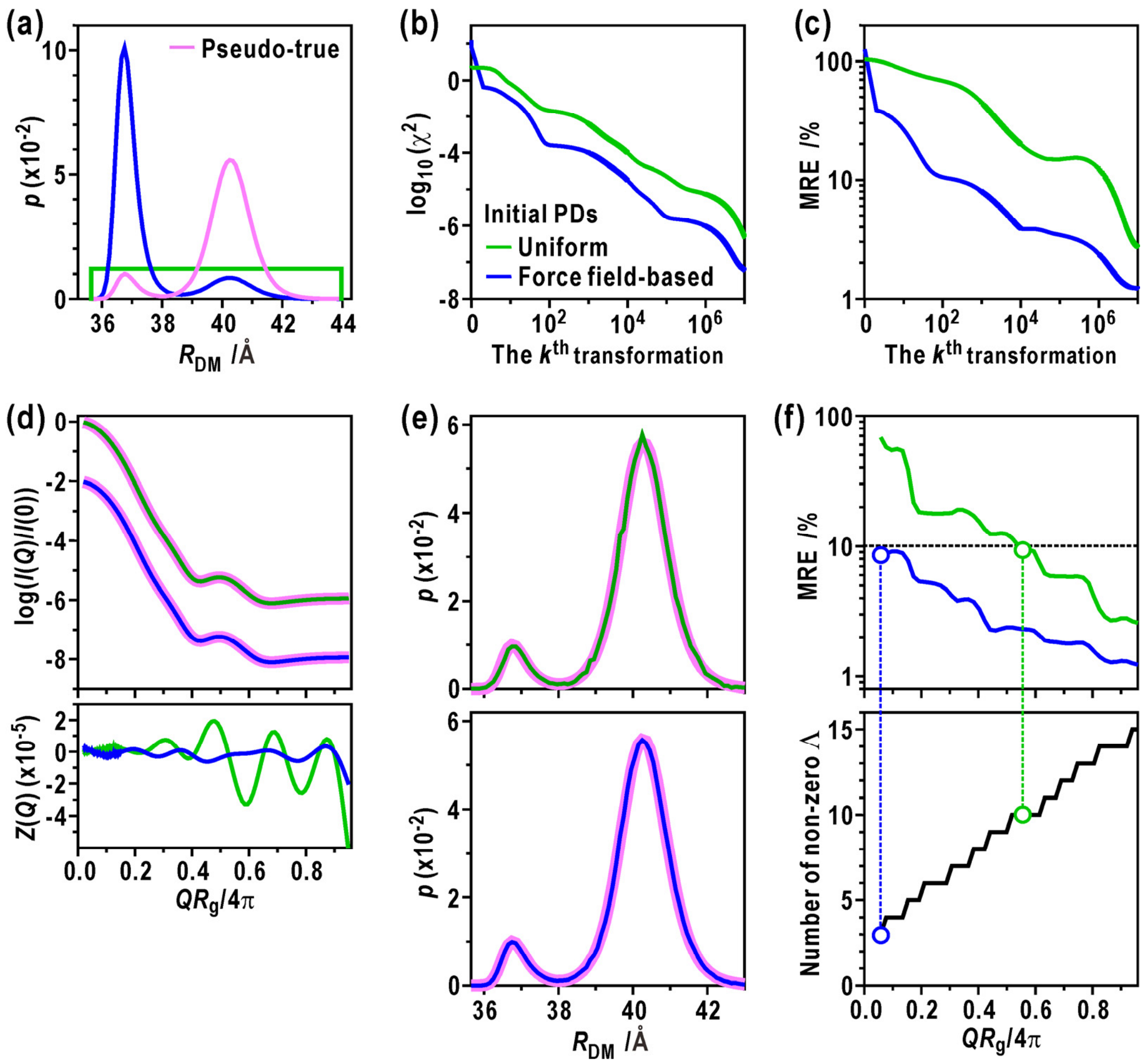

**FIG. S6.** Results of the simulations reconstructing conformational ensemble of *Mt*EPSPS from its pseudo-experimental SAXS data using the MBGD method. The green and blue represent the results for MBGD, which were initiated from the uniform and force field-based PDs. (a) Protein conformational ensembles were represented as PDs on $R_{\mathrm{DM}}$ in the case of *Mt*EPSPS, showing the pseudo-true (pink), uniform (green), and force field-based (blue) PDs. (b-c) Minimization processes monitored through $\chi^2$ (b) and AREP (c). (d) Comparisons of the pseudo-experimental SAXS data (red) and those obtained using MBGD. Residuals between the pseudo-experimental and calculated data were evaluated using $z(Q)$ (lower). (e) Comparisons of the pseudo-true PD and those reconstructed using MBGD. (f) Dependencies of the reconstruction accuracy monitored using AREP (upper) and the number of nonzero eigenvalues (lower) on the amount of data. The white-filled circles represent the upper bound of the data range. Narrowing the range further results in an AREP exceeding 10% (black dashed line).

Using the generated *Mt*EPSPS ensembles, we conducted the reconstruction simulations utilizing the MBGD method. The SAXS data calculated from the pseudo-true PD were utilized as the pseudo-experimental data, with the $\chi^2$ function serving as the objective functional. The initial PDs were the uniform and force field-based PDs (Fig. S6(a)). The MBGD calculation succeeded in both reducing the $\chi^2$ value (Fig. S6(b) and S6(d)) and reconstructing the pseudo-true PD (Fig. S6(c) and S6(e)) when employing the pseudo-experimental SAXS data up to $QR_g/4\pi = 0.96$. Initiating the MBGD calculation from the uniform PD yielded a pseudo-true PD reconstruction with an accuracy of AREP = 3% (upper panel in Fig. S6(e)). Initiating the calculation from the force field-based PD led to near-complete reconstruction with an accuracy of AREP = 1% (the lower panel in Fig. S6(e)). The numbers of nonzero eigenvalues of the Hessian of the objective functional necessary to achieve reconstruction accuracy exceeding AREP = 10% when initiating from the uniform and force field-based PDs were 10 and 3, respectively (Fig. S6(f)).

**Note S6. Reconstruction simulation of conformational ensembles of guanylate kinase from SAXS data**

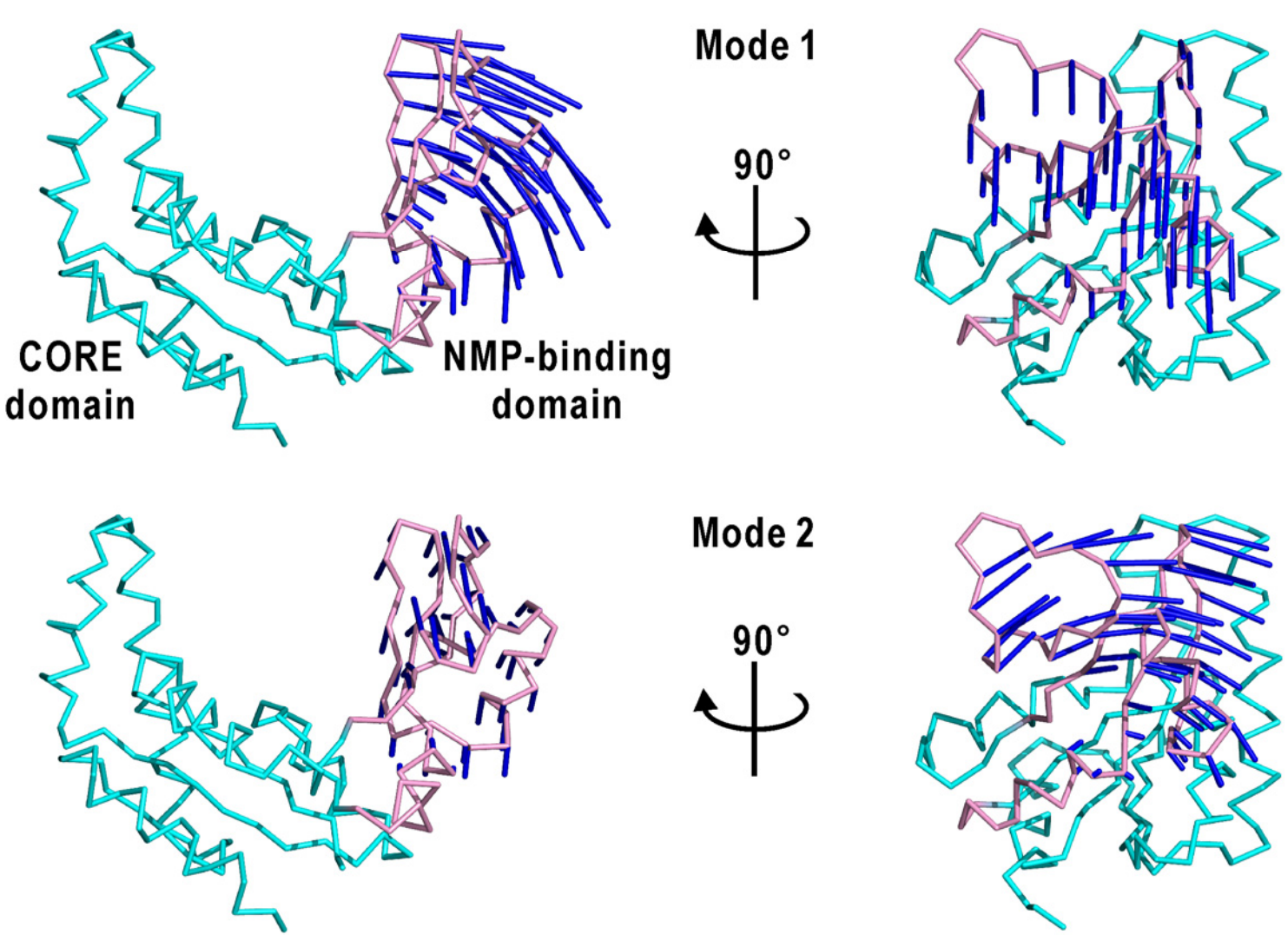

**FIG. S7.** Conformational motions of GK. PCA modes calculated from CGMD trajectories: first PC mode (upper) and second PC mode (lower). The amplitude ($3\sigma$) and direction of $C_\alpha$ displacements along the first PC mode are represented by blue arrows. The domain structures are represented using different colors.

As the third model system for the validation test of the MBGD method, we employed guanylate kinase (GK) [23]. The structure of GK comprises two domains: CORE and NMP-binding domains (Fig. S7(a)). The crystal structures revealed that substrate binding induces domain-closure motion between these domains. We conducted the CGMD simulation using the crystal structure of the apo form (PDB ID:1EX6) for the Gō model potential. To characterize the conformational motions of GK in the CGMD simulation, we analyzed the trajectory through PCA. Most of the total $C^{\alpha}$-mean square fluctuations were accounted for by the first and second PC modes (38 and 24%, respectively). These modes represented domain motions, which were open-close and twisting motions of the two domains, respectively (Fig. S7).

The model PDs of GK for the reconstruction simulations were generated through the multiple-Gō CGMD simulation (Supplementary Methods). The crystal structures in the holo form (PDB ID: 1GKY [24]) and the apo form served as inputs for the metastable open and closed conformational states, respectively. Adjusting parameters in the multiple-Gō model potential produced two broad ensembles, primarily populated in closed and open conformational states (Fig. S8(a) and S8(b), respectively). These ensembles were used as the force field-based and pseudo-true PDs, respectively.

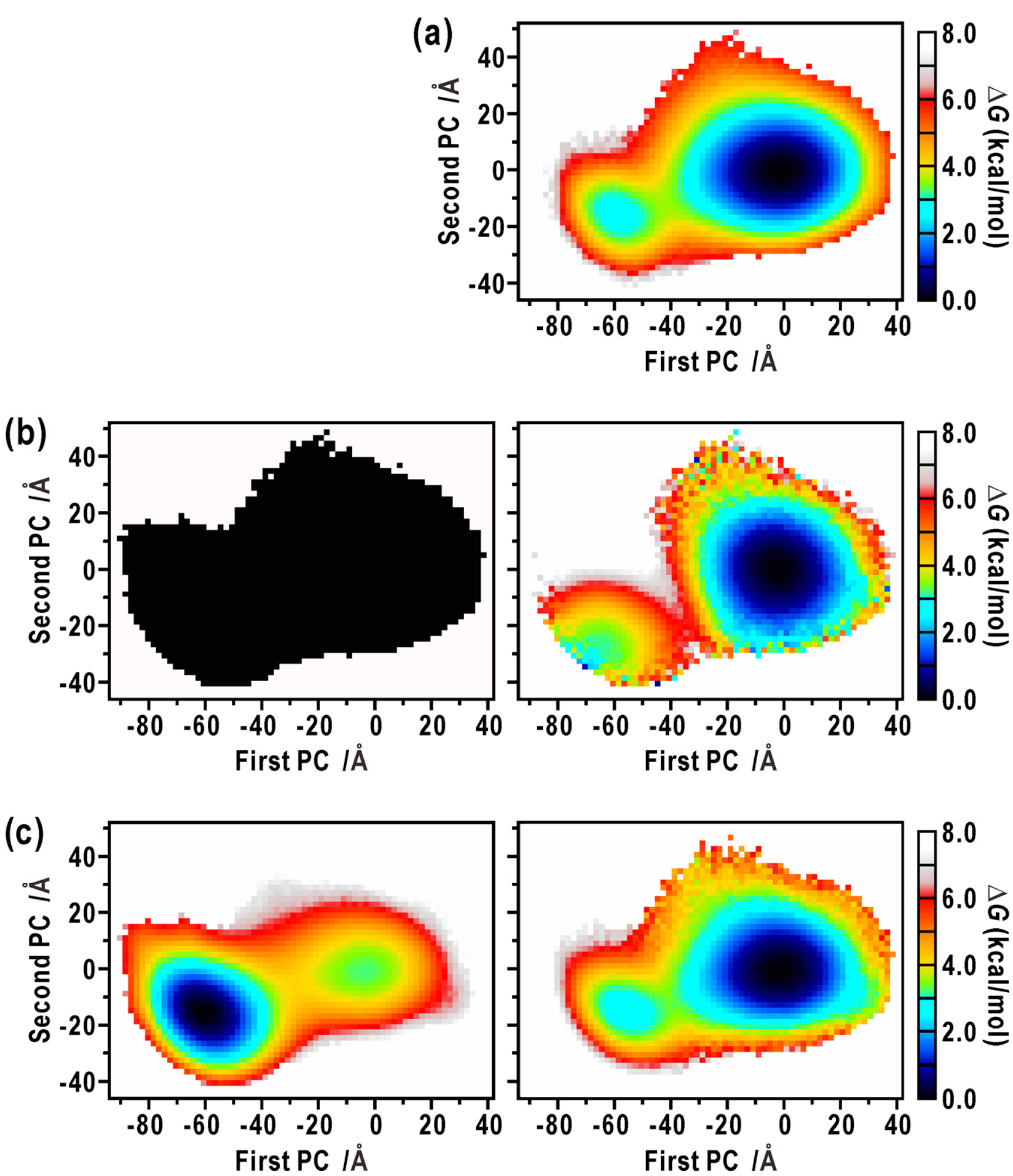

**FIG. S8.** Protein conformational ensembles were represented as PDs on the first and second PC modes in the case of GK. (a) The pseudo-true PD. (b) The uniform PD as an initial for the reconstruction simulation (left). The right panel displays the PD reconstructed by the MBGD method. (c) The force field-based PD as an initial for the reconstruction simulation (left). The right panel displays the PD reconstructed by the MBGD method.

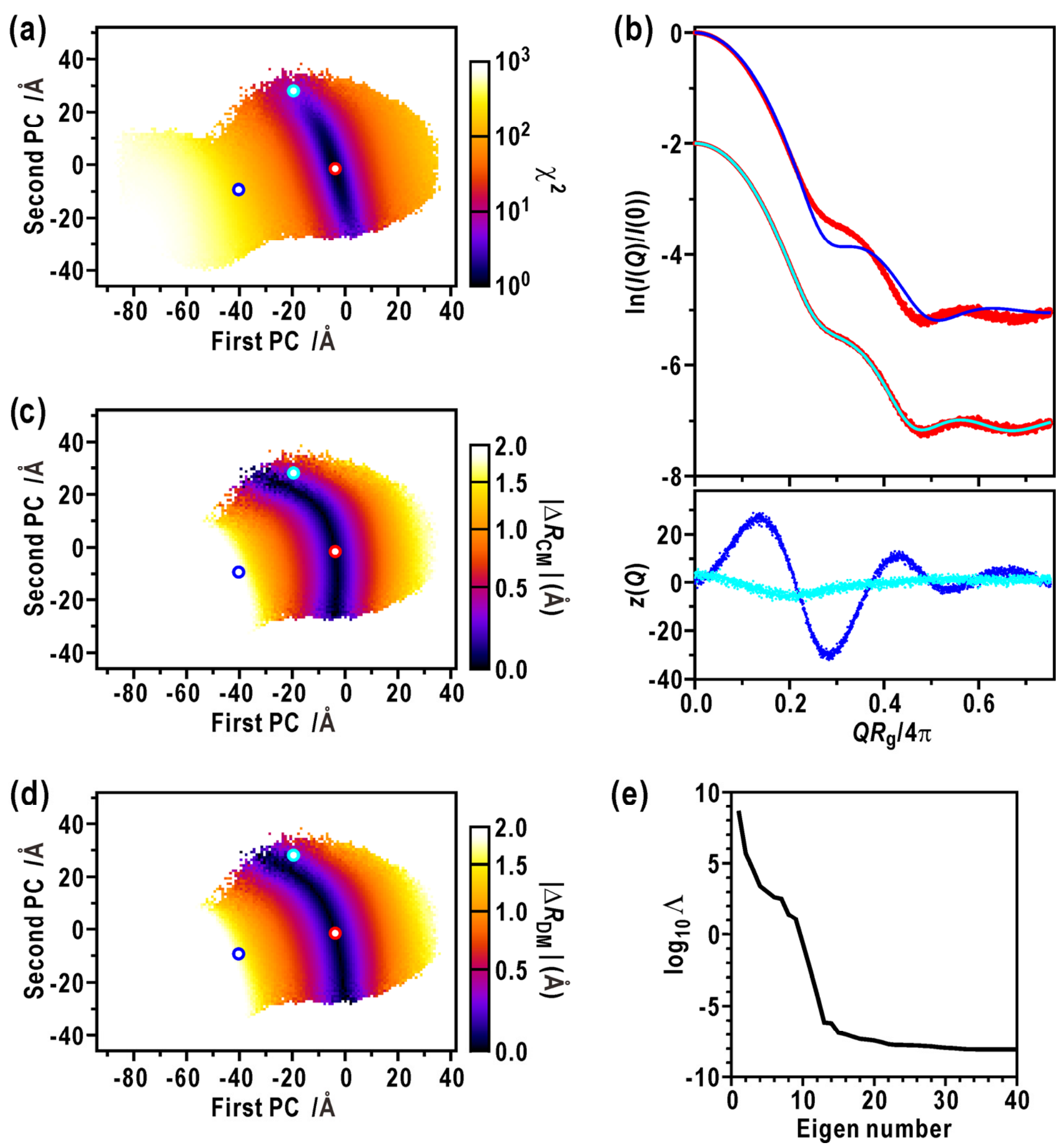

**FIG. S9.** Dependencies of the calculated SAXS data on conformational changes of GK. (a) Dependencies of $\chi^2$ values on domain motions outlined by the first and second PC modes. The red circle indicates the conformation from which the pseudo-experimental SAXS data were generated. (b) Comparison of the pseudo-experimental SAXS data (red) with the data calculated from the two representative conformations, representing those indicated by the blue and cyan circles in (a). Residuals between the data were evaluated using $z(Q)$ (lower). (c) Dependencies of the coordinate $R_{\mathrm{CM}}$ on the domain motions outlined by the first and second PC modes. The absolute value of the difference in $R_{\mathrm{CM}}$, $|\Delta R_{\mathrm{CM}}|$, from the conformation utilized to generate pseudo-experimental SAXS data (red circle) was plotted by color. (d) Dependencies of the coordinate $R_{\mathrm{DM}}$ on the domain motions outlined by the first and second PC modes. The absolute value of the difference in $R_{\mathrm{DM}}$, $|\Delta R_{\mathrm{DM}}|$, from the conformation utilized to generate pseudo-experimental SAXS data (red circle) was plotted by color. (f) Distribution of the eigenvalues of the Hessian of the objective functional when utilizing the first and second PC modes to describe the conformational motions.

To select the coordinates on which PDs are projected, we investigated the dependence of the calculated SAXS data on the conformational changes of GK. To describe the conformational changes, we used the PC modes obtained from the PCA conducted on the CGMD simulation mentioned previously. Among the PC modes, the calculated SAXS data showed dependencies only for the first two PC modes, which were domain motions. Therefore, we observed the dependence of the data on the 2D plane spanned by the first and second PC modes (Fig. S9(a)). All conformations within a 1 Å grid on the plane were categorized as a single conformational state represented by this grid. To observe the SAXS data dependencies, we calculated the $\chi^2$ values for all pairs of conformational states following the same procedure used for transferrin (Note S4).

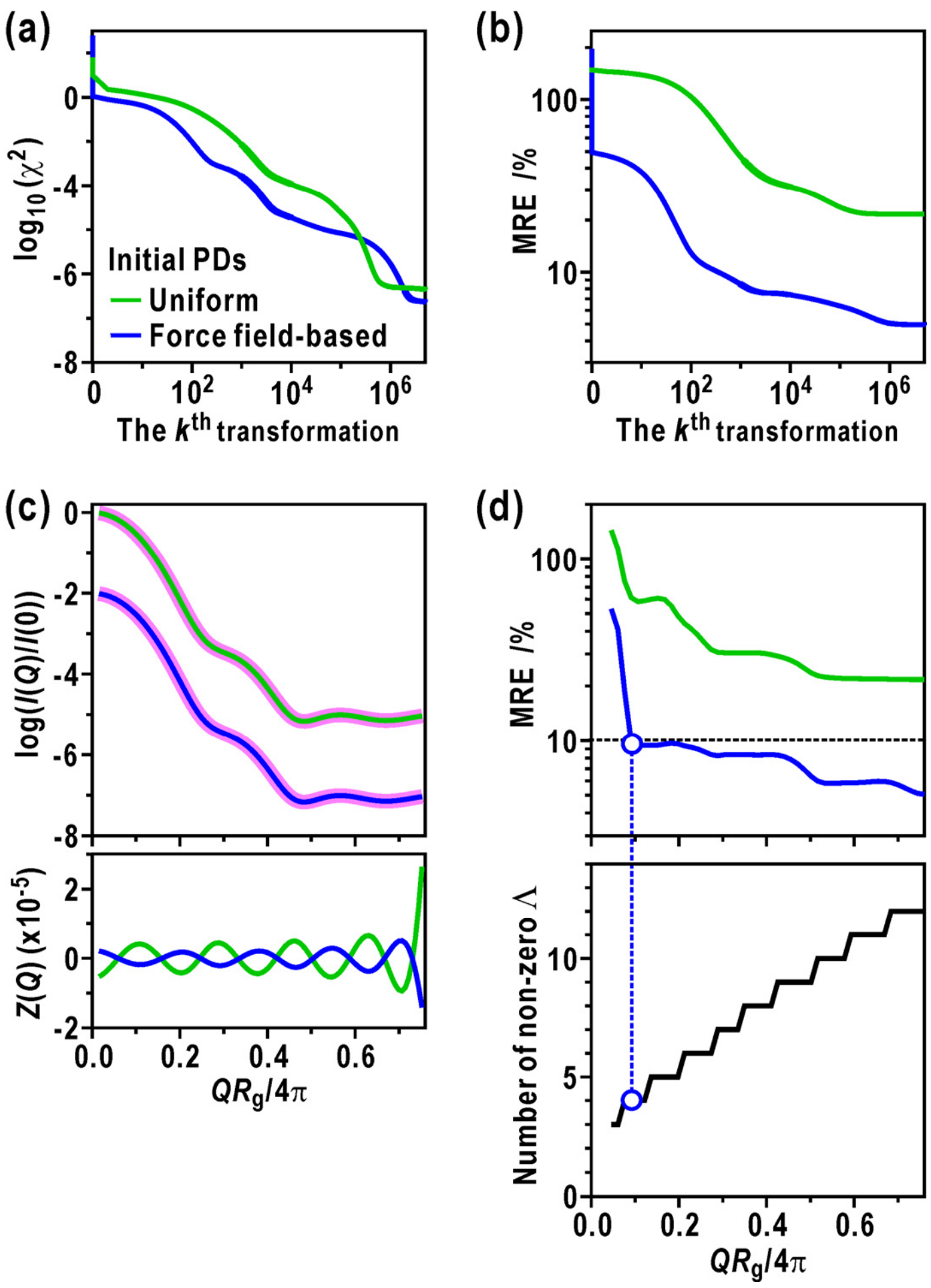

**FIG. S10.** Results of the simulations reconstructing conformational ensemble of GK from its pseudo-experimental SAXS data using the MBGD method. The green and blue represent the results for MBGD, which were initiated from the uniform and force field-based PDs. (a-b) Minimization processes monitored through $\chi^2$ (a) and AREP (b). (c) Comparisons of the pseudo-experimental SAXS data (red) and those obtained using MBGD. Residuals between the pseudo-experimental and calculated data were evaluated using $z(Q)$ (lower). (d) Dependencies of the reconstruction accuracy monitored using AREP (upper) and the number of nonzero eigenvalues (lower) on the amount of data. The white-filled circle represent the upper bound of the data range. The black dashed line represents an AREP exceeding 10%.

Fig. S9(a) shows an example of the $\chi^2$ plot on the 2D plane, which was calculated using the reference conformational state indicated by the red circle as the state providing the pseudo-experimental data. As shown in this plot, the SAXS data depended on conformational changes along the first and second PC modes. There is a banded-shape region where the $\chi^2$ value from the reference conformational state remains small (Fig. S9(a) and S9(b)). However, unlike transferrin and *Mt*EPSPS, there are certain differences in SAXS data between the reference and the conformational state located at the edge of that region. Additionally, in both reaction coordinates $R_{CM}$ and $R_{DM}$, the regions where the values of these coordinates remain the same do not coincide with the above region (Fig. S9(c) and S9(d), respectively). We could not find another reaction coordinate satisfying this coincidence. Therefore, we decided to use the first and second PC modes as the 2D coordinate systems to describe the variations in the SAXS data. When utilizing these PC modes and SAXS data up to $QR_g/4\pi = 0.76$ ($Q = 0.5$ Å$^{-1}$), the number of nonzero eigenvalues of the Hessian of the $\chi^2$ function was 12 (Fig. S9(e)).

Using the generated GK ensembles, we conducted the reconstruction simulations utilizing the MBGD method. The SAXS data calculated from the pseudo-true PD (Fig. S8(a)) were utilized as the pseudo-experimental data, with the $\chi^2$ function serving as the objective functional. The initial PDs were the uniform and force field-based PDs (The left panels in Figs. S8(b) and S8(c), respectively). When employing the force field-based PD as an initial with the pseudo-experimental SAXS data up to $QR_g/4\pi = 0.76$, the MBGD calculation succeeded in both reducing the $\chi^2$ value (Figs. S10(a) and S10(c)) and reconstructing the pseudo-true PD (the right panel in Fig. S8(c), and Fig. S10(b)). The accuracy of the reconstructed PD was AREP = 5%. However, initiating the MBGD calculation from the uniform PD yielded a reconstructed PD with an accuracy of AREP = 21% (the right panel in Fig. S8(b)), even though the $\chi^2$ value was reduced. The numbers of nonzero eigenvalues of the Hessian of the objective functional necessary to achieve reconstruction accuracy exceeding AREP = 10% when initiating from the force field-based PDs were 4 (Fig. S10(d)).

**Note S7. Details in application of MBGD to experimental SAXS data of SjGlcNK**

To select the coordinates on which PDs are projected, we investigated the dependence of the calculated SAXS data on the conformational changes of SjGlcNK. To describe the conformational changes, we used the first and second PC modes obtained from the PCA conducted on the AA+CGMD trajectories. Among the PC modes, the calculated SAXS data showed dependencies only for the first two PC modes, which were domain motions. Therefore, we observed the dependence of the data on the 2D plane spanned by the first and second PC modes (Fig. S9(a)). All conformations within a 2 Å grid on the plane were categorized as a single conformational state represented by this grid. To observe the SAXS data dependencies, we calculated the $\chi^2$ values for all pairs of conformational states following the same procedure used for transferrin (Note S4).

Fig. S11(a) shows an example of the $\chi^2$ plot on the 2D plane, which was calculated using the reference conformational state indicated by the red circle as the state providing the pseudo-experimental data. As shown in this plot, the SAXS data depended on conformational changes along the first and second PC modes. There is a banded-shape region where the $\chi^2$ value from the reference conformational state remains small (Fig. S11(a) and S11(b)). In both reaction coordinates $R_{CM}$ and $R_{DM}$, the regions where the values of these coordinates remain the same do not coincide with the above region (Fig. S11(c) and S11(d), respectively). Therefore, we decided to use the first and second PC modes as the 2D coordinate systems to describe the variations in the SAXS data. When utilizing these PC modes and SAXS data up to $QR_{\mathrm{g}}/4\pi = 0.41$ ($Q = 0.2$ Å$^{-1}$), the number of nonzero eigenvalues of the Hessian of the $\chi^2$ function was 9 (Fig. S11(e)).

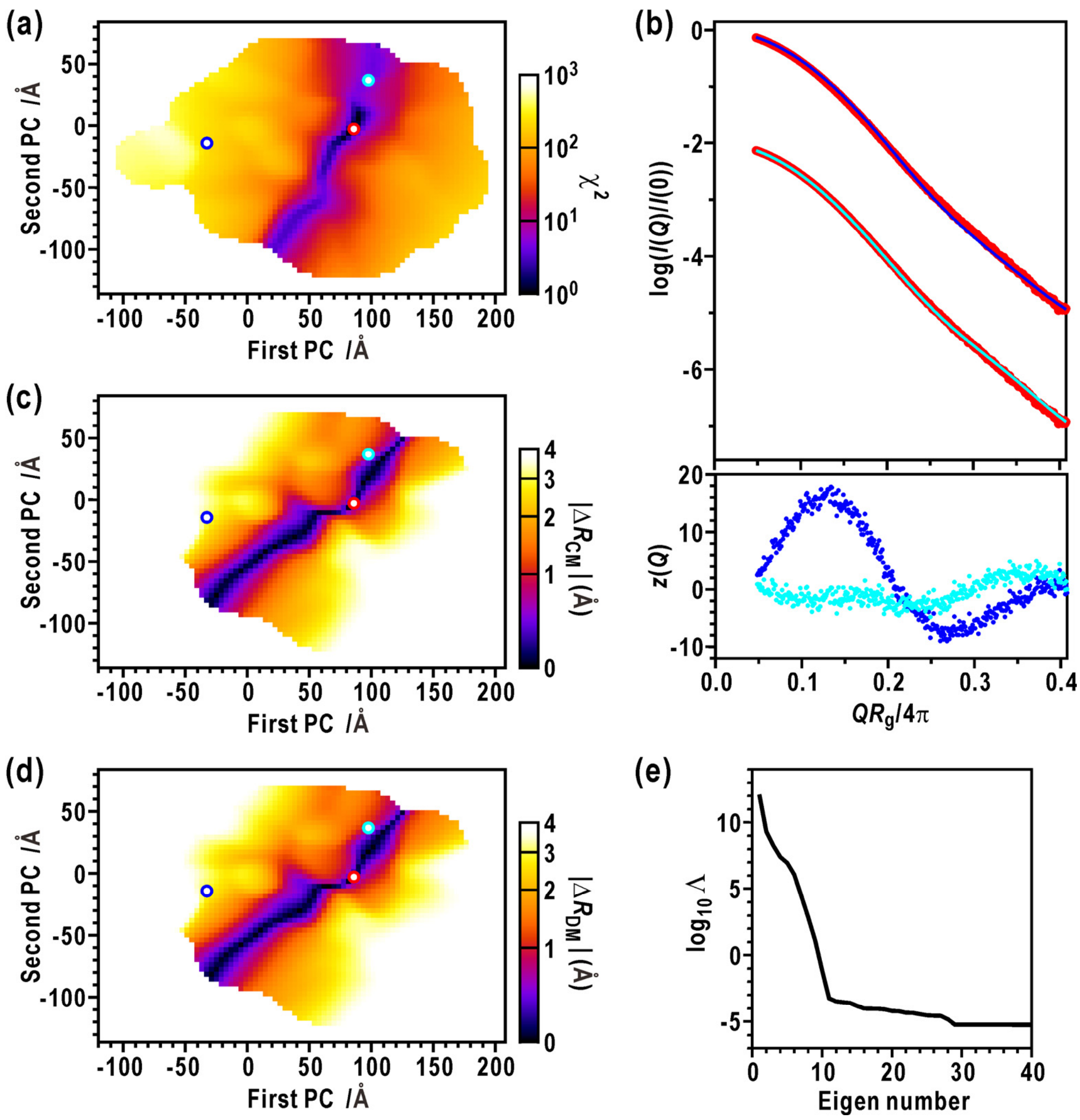

**FIG. S11.** Dependencies of the calculated SAXS data on conformational changes of SjGlcNK. (a) Dependencies of $\chi^2$ values on domain motions outlined by the first and second PC modes. The red circle indicates the conformation from which the pseudo-experimental SAXS data were generated. (b) Comparison of the pseudo-experimental SAXS data (red) with the data calculated from the two representative conformations, representing those indicated by the blue and cyan circles in (a). Residuals between the data were evaluated using $z(Q)$ (lower). (c) Dependencies of the coordinate $R_{\mathrm{CM}}$ on the domain motions outlined by the first and second PC modes. The absolute value of the difference in $R_{\mathrm{CM}}$, $|\Delta R_{\mathrm{CM}}|$, from the conformation utilized to generate pseudo-experimental SAXS data (red circle) was plotted by color. (d) Dependencies of the coordinate $R_{\mathrm{DM}}$ on the domain motions outlined by the first and second PC modes. The absolute value of the difference in $R_{\mathrm{DM}}$, $|\Delta R_{\mathrm{DM}}|$, from the conformation utilized to generate pseudo-experimental SAXS data (red circle) was plotted by color. (f) Distribution of the eigenvalues of the Hessian of the objective functional when utilizing the first and second PC modes to describe the conformational motions.

In the present MBGD application, we used experimental SAXS data acquired under substrate-free conditions [14] (Fig. 8(c)). In practice, SAXS data typically include unknown experimental parameters such as intensity scaling factor and solvation density [9,15]. We incorporated a procedure to estimate these parameters into the MBGD algorithm. Prior to the application, we also examined the impact of two important factors on the reconstruction of conformational ensembles: (i) noise present in experimental data, and (ii) insufficient sampling in MD simulations used to generate model ensembles as initial inputs. The reconstructions obtained using MBGD demonstrated excellent robustness against noise levels typically observed in SAXS database [10]. In addition, we found that insufficient sampling in MD simulations resulted in reconstructed ensembles exhibiting density discontinuities at the edges, which are therefore distinguishable from those generated with sufficient sampling. While the procedures to estimate the unknown experimental parameters and to identify insufficient sampling in the MBGD method are challenging in themselves, they are domain-specific and therefore will be reported separately. In the main text, we focus on the results obtained by applying MBGD to the actual experimental data.

We first conducted the MBGD calculation using the AAMD-derived ensemble as the initial input. However, this approach resulted in minimization failure and a reconstructed ensemble with density discontinuities at its edges, owing to insufficient conformational sampling by AAMD (Fig. S12). Furthermore, we also conducted the MBGD calculation starting from an ensemble that included only the equilibrium fluctuations around the crystal structure conformations. The results showed a reconstructed ensemble with discontinuities (Fig. S13), demonstrating that the experimental data could not be considered without dropped-jar conformations.

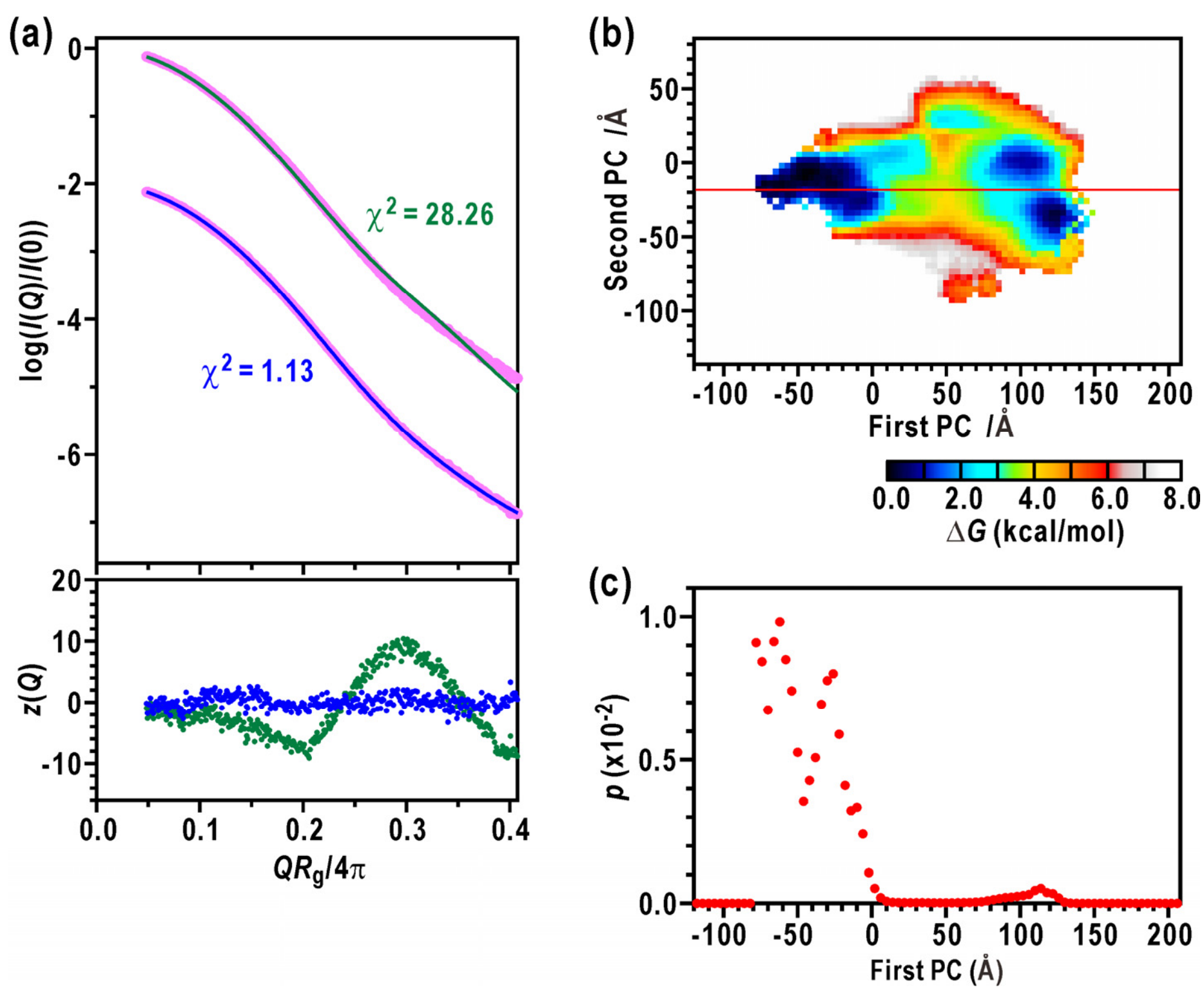

**FIG. S12.** Results of the simulations reconstructing conformational ensemble of SjGlcNK from its experimental SAXS data using the MBGD method. The AAMD-derived ensemble was employed as an initial. (a) Comparison of the experimental (pink) and calculated SAXS data. The green and blue curves represent the data calculated from the AAMD and reconstructed ensembles, respectively. (b) The MBGD-reconstructed ensemble projected on the first and second PC modes. (c) Plot of the PD along the red line in the PMF map (b).

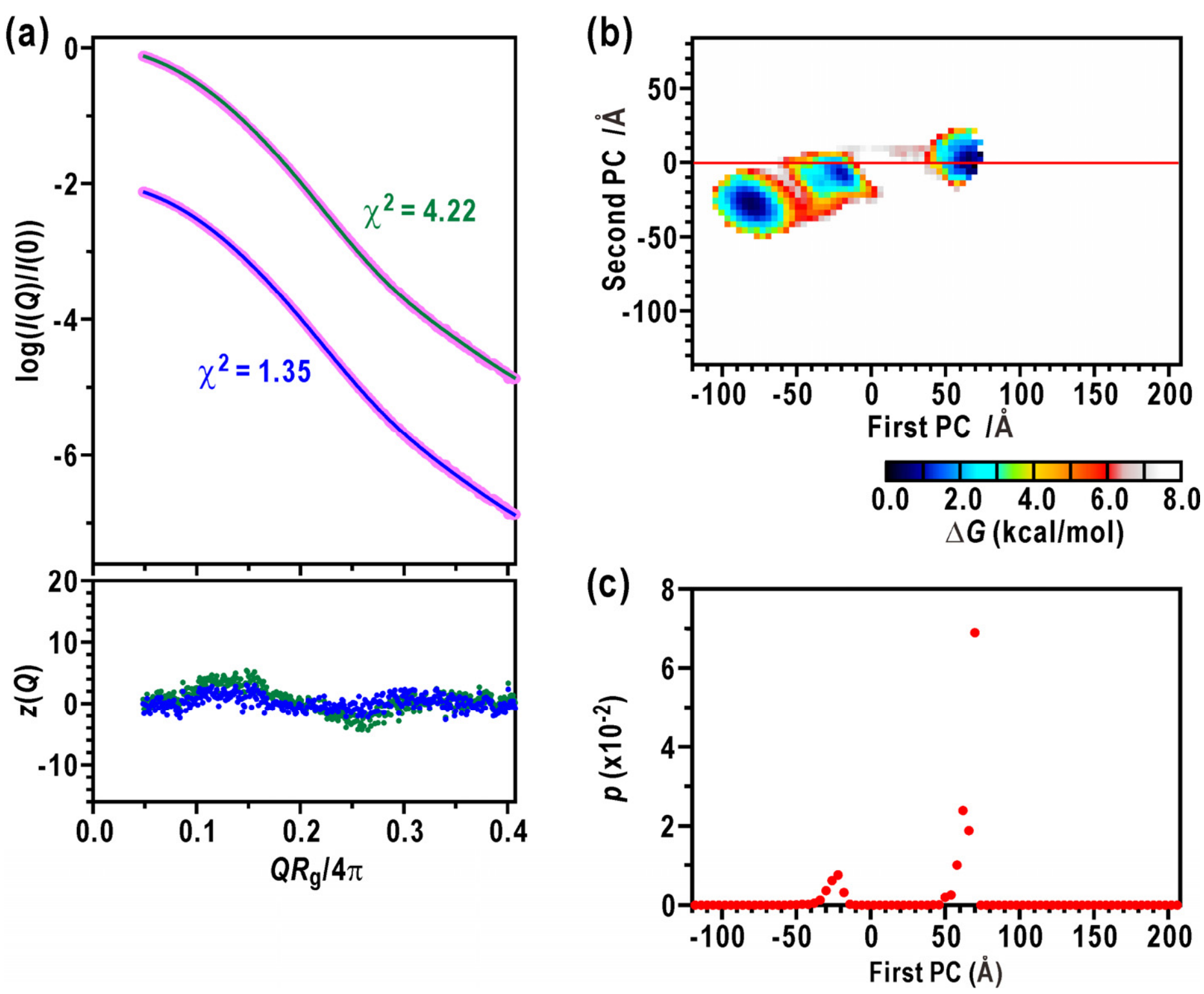

**FIG. S13.** Results of the simulations reconstructing conformational ensemble of SjGlcNK from its experimental SAXS data using the MBGD method. The reconstruction was initiated from an ensemble that included only the equilibrium fluctuations around the crystal structure conformations. (a) Comparison of the experimental (pink) and calculated SAXS data. The green and blue curves represent the data calculated from the initial and reconstructed ensembles, respectively. (b) The MBGD-reconstructed ensemble projected on the first and second PC modes. (c) Plot of the PD along the red line in the PMF map (b).